\documentclass[%
 reprint,
superscriptaddress,
 amsmath,amssymb,
 aps,
]{revtex4-2}
\UseRawInputEncoding
\usepackage{graphicx}
\usepackage{dcolumn}
\usepackage{bm}
\usepackage{float}

\begin{document}

\preprint{APS/123-QED}

\title{Magnetic tuning of the tunnel coupling in an optically active quantum dot molecule}

\author{Frederik Bopp}
\altaffiliation{These authors contributed equally to this work}

\affiliation{%
 Walter Schottky Institut, School of Natural Sciences, and MCQST, Technische Universit\"at M\"unchen, Am Coulombwall 4, 85748 Garching, Germany
}%

\author{Charlotte Cullip}
\altaffiliation{These authors contributed equally to this work}
\affiliation{%
 Walter Schottky Institut, School of Natural Sciences, and MCQST, Technische Universit\"at M\"unchen, Am Coulombwall 4, 85748 Garching, Germany
}%

\author{Christopher Thalacker}
\affiliation{%
 Walter Schottky Institut, School of Natural Sciences, and MCQST, Technische Universit\"at M\"unchen, Am Coulombwall 4, 85748 Garching, Germany
}%

\author{Michelle Lienhart}
\affiliation{%
 Walter Schottky Institut, School of Natural Sciences, and MCQST, Technische Universit\"at M\"unchen, Am Coulombwall 4, 85748 Garching, Germany
}%

\author{Johannes Schall}
\affiliation{%
 Institut f\"ur Festk\"orperphysik, Technische Universit\"at Berlin, Hardenbergstraße 36, 10623 Berlin, Germany
}%

\author{Nikolai Bart}
\affiliation{%
 Lehrstuhl f\"ur Angewandte Festk\"orperphysik, Ruhr-Universit\"at Bochum, Universit\"atsstraße 150, 44801 Bochum, Germany
}%

\author{Friedrich Sbresny}
\affiliation{%
 Walter Schottky Institut, School of Computation, Information and Technology, and MCQST, Technische Universit\"at M\"unchen, Am Coulombwall 4, 85748 Garching, Germany
}%

\author{Katarina Boos}
\affiliation{%
 Walter Schottky Institut, School of Computation, Information and Technology, and MCQST, Technische Universit\"at M\"unchen, Am Coulombwall 4, 85748 Garching, Germany
}%

\author{Sven Rodt}
\affiliation{%
 Institut f\"ur Festk\"orperphysik, Technische Universit\"at Berlin, Hardenbergstraße 36, 10623 Berlin, Germany
}%

\author{Dirk Reuter}
\affiliation{%
Paderborn University, Department of Physics, Warburger Straße 100, 33098 Paderborn, Germany
}%

\author{Arne Ludwig}
\affiliation{%
 Lehrstuhl f\"ur Angewandte Festk\"orperphysik, Ruhr-Universit\"at Bochum, Universit\"atsstraße 150, 44801 Bochum, Germany
}%

\author{Andreas D. Wieck}
\affiliation{%
 Lehrstuhl f\"ur Angewandte Festk\"orperphysik, Ruhr-Universit\"at Bochum, Universit\"atsstraße 150, 44801 Bochum, Germany
}%

\author{Stephan Reitzenstein}
\affiliation{%
 Institut f\"ur Festk\"orperphysik, Technische Universit\"at Berlin, Hardenbergstraße 36, 10623 Berlin, Germany
}%

\author{Filippo Troiani}
\affiliation{%
 Centro S3, CNR-Istituto Nanoscienze, Via Campi 213/a, 41125 Modena, Italy
}%

\author{Guido Goldoni}
\affiliation{%
Centro S3, CNR-Istituto Nanoscienze, Via Campi 213/a, 41125 Modena, Italy
}%
\affiliation{%
Dipartimento di Scienze Fisiche, Informatiche e Matematiche, Universit\`a di Modena e Reggio Emilia, Via Campi 213/a, 41125 Modena, Italy
}%

\author{Elisa Molinari}
\affiliation{%
Centro S3, CNR-Istituto Nanoscienze, Via Campi 213/a, 41125 Modena, Italy
}%
\affiliation{%
Dipartimento di Scienze Fisiche, Informatiche e Matematiche, Universit\`a di Modena e Reggio Emilia, Via Campi 213/a, 41125 Modena, Italy
}%

\author{Kai M\"uller}
\affiliation{%
 Walter Schottky Institut, School of Computation, Information and Technology, and MCQST, Technische Universit\"at M\"unchen, Am Coulombwall 4, 85748 Garching, Germany
}%

\author{Jonathan J. Finley}%
 \email{finley@wsi.tum.de}
\affiliation{%
 Walter Schottky Institut, School of Natural Sciences, and MCQST, Technische Universit\"at M\"unchen, Am Coulombwall 4, 85748 Garching, Germany
}%

\date{\today}

\begin{abstract}
Self-assembled optically active quantum dot molecules (QDMs) allow the creation of protected qubits via singlet-triplet spin states. The qubit energy splitting of these states is defined by the tunnel coupling strength and is, therefore, determined by the potential landscape and thus fixed during growth. Applying an in-plane magnetic field increases the confinement of the hybridized wave functions within the quantum dots, leading to a decrease of the tunnel coupling strength. We achieve a tuning of the coupling strength by $(53.4\pm1.7)$\,\%. The ability to fine-tune this coupling is essential for quantum network and computing applications that require quantum systems with near identical performance.


\end{abstract}

\maketitle




Although any physical qubit is susceptible to dephasing, the use of protected qubits offers a way to passively shield the quantum information from its noise environment\,\cite{Lidar1998,Doucot2012}. Few-spin solid-state qubits are particularly promising in this regard, since they are immune to the dominant electric and magnetic field noise sources to first order\,\cite{Petta2005,Stopa2008,Weiss2012}.
Recently, the extension of the spin coherence time by creating a few-spin (e.g., singlet-triplet) qubit has been demonstrated in a vertically stacked pair of optically active quantum dots\,\cite{Tran2022}. Such quantum dot molecules (QDMs) combine the advantages of single quantum dots (QDs), such as robust optical selection rules \cite{Benny2011} and dominant emission into the zero-phonon line \cite{Favero2003}, with the ability to create protected few-spin qubits \cite{Weiss2012}. They are formed from two vertically stacked QDs separated by a tunneling barrier. At a separation of a few nanometers, the orbital states in the two dots become tunnel coupled with a strength that depends on the separation of the dots and the height of the tunnel barrier\,\cite{Bracker2006}. Tunnel coupling facilitates the creation of hybridized symmetric and anti-symmetric few-spin states, which serve as protected qubit eigenstates\,\cite{Martins2016}. The energy splitting between these two states is defined by the tunnel coupling. This makes the coupling strength the main tuning parameter of the solid-state qubit, since it determines the operating rate of quantum gates\,\cite{Kim2011}. However, the thickness and height of the tunnel barrier are fixed during the growth process. By establishing a method for post-growth continuous and reversible tuning of the tunnel coupling, synchronization of multiple qubits could be enabled to facilitate applications such as quantum networks\,\cite{Kimble2008}. In addition, the generation of two-dimensional photonic cluster states for measurement-based quantum computing could be optimized\,\cite{Economou2010, Raussendorf2001}.

\begin{figure}
\includegraphics{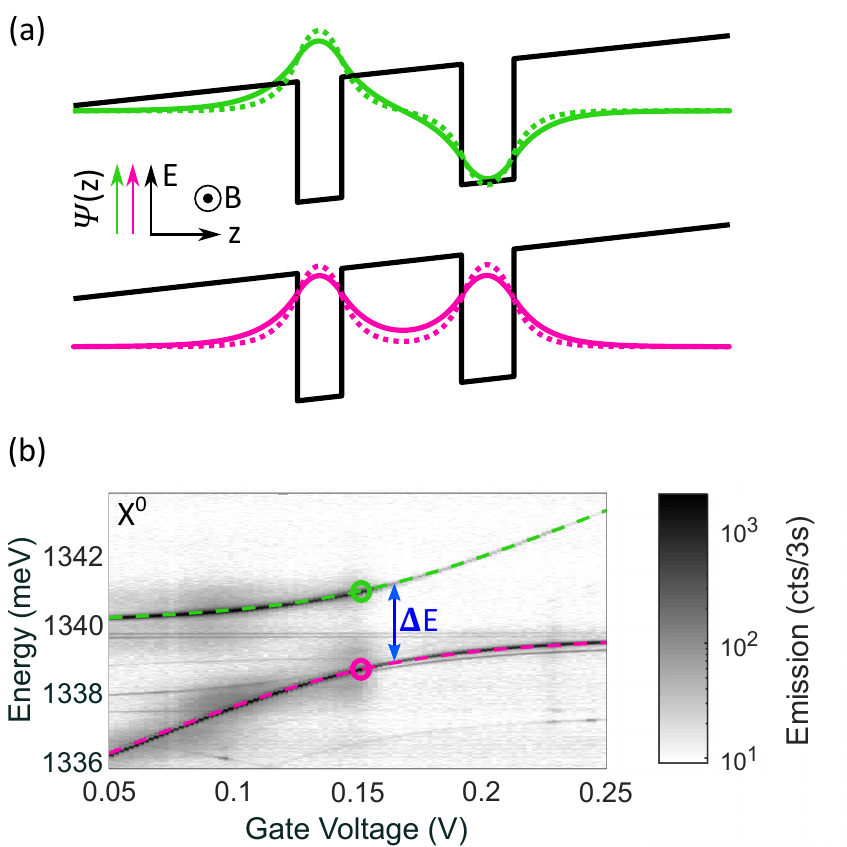}
\caption{\label{fig:1}Tunnel coupling of the neutral exciton ($X^0$). (a) Schematic of bonding (pink) and anti-bonding (green) wave function $\Psi(z)$ of the lowest and second lowest electron energy eigenstate, respectively, with (dashed) and without (solid) applied in-plane magnetic field $B$. The solid black lines depict the electron trapping potential along the growth direction $z$. (b) Voltage-dependent photoluminescence measurement of the $X^0$. The energy splitting between the bonding (pink) and anti-bonding (green) eigenstates $\Delta E$ is depicted by a blue arrow, while the energy of the tunnel-coupled states is visualized by colored circles.}
\end{figure}

One approach to control the tunnel coupling post-growth is to apply strain to the system. However, so far only weak energy shifts up to 50\,$\mu$eV have been achieved experimentally \cite{Zallo2014} while lacking long-term stability\,\cite{Carter2019}. A second approach to tune the tunnel coupling, promising a wider tuning range with higher temporal stability, is to apply a static magnetic field to the QDM. This results in a modification of the orbital part of the carrier wave function and, therefore, control of the tunnel coupling.
Over the past decades, several theoretical works have explored the magnetic tuning of symmetric and anti-symmetric orbitals in artificial molecules\, \cite{Burkard2000,Korkusinski2001,Bellucci2004,Jacob2004,DiasDaSilva2007,Climente2008}.
However, experimental demonstrations have not yet been provided\,\cite{Planelles2010,Phoenix2022} and structure-property relationships are yet to be established. In-plane magnetic fields are predicted to result in a confinement of the orbital part of the wave function along a direction perpendicular to the field direction (diamagnetic response), corresponding to a reduction of the tunnel coupling\,\cite{Bellucci2004}.
To qualitatively illustrate the magnetic field dependence of the wave function, Figure\,\ref{fig:1}\,(a) shows a sketch of the electron wave function in a double-well potential without (solid) and with (dashed) magnetic field $B$ applied along the x-direction. The pink (green) wave function $\Psi(z)$ illustrates the lowest symmetric (second lowest, anti-symmetric) eigenstate of a single electron confined in the QDM potential. The solid black lines represent an approximate potential landscape of the QDM along the growth direction $z$. The probability density of the charge position inside the two potential wells is predicted to increase at higher magnetic field due to magnetic compression of the wave function\,\cite{Bellucci2004}. Consequently, the wave function is expelled from the barrier region between the two potentials wells, which effectively reduces the strength of the tunnel coupling.

In this work, we experimentally demonstrate the tunability of the tunnel coupling of an artificial molecule by varying an in-plane magnetic field. We perform magneto-photoluminescence measurements on a single QDM to study the energy splitting between the bonding (BO) and anti-bonding (AB) neutral exciton ($X^0$) states and hence the tunnel coupling strength. We observe a reduced splitting with increasing magnetic field, which we numerically simulate by modeling a three-dimensional confinement potential and calculating the corresponding single-particle energies and Coulomb interaction between the localized carriers. By applying an electric field along the growth direction, we tune the electronic levels of the two dots into and out of resonance. This allows us to conclusively show that the magnetic field leads to a reduction of the tunnel coupling strength, providing a post-growth control of this parameter.


The QDM investigated in this work consists of two vertically stacked and tunnel-coupled InAs QDs embedded in a GaAs matrix\,\cite{Krenner2005}. The neutral exciton ($X^0$) is used to demonstrate the general reduction of the energy gap between BO and AB eigenstates. Excitation of the neutral exciton is achieved via a continuous wave laser resonant to a higher orbital shell at 1353.6\,meV, while the s-shell emission is monitored. To efficiently excite the $X^0$ state, a two-phase optical and electrical sequence is used for all measurements presented in this letter\,\cite{Bopp2022}. The applied gate voltage allows tuning of the energy levels of the two QDs relative to each other. In combination with the tunnel coupling, this allows the hybridization of charge wave functions forming BO and AB eigenstates. An avoided crossing is the result.
Figure\,\ref{fig:1}\,(b) shows the voltage-dependent photoluminescence of the $X^0$ state. The eigenenergies corresponding to the BO and AB eigenstates are highlighted by a dashed pink and green fit, respectively.
The exciton energies are fitted using a coupled two-state model, including a linear bias and a quadratic contribution proportional to the bias, which accounts for the DC Stark-effect for indirect and direct excitons, respectively\,\cite{Fry2000a}. The full description can be found in the Supplemental Material\,\cite{Supp}.

The voltage at which hybridization of the two wave functions occurs is depicted by colored circles in Figure\,\ref{fig:1}\,(b). The minimum energy difference $\Delta E$ between the two exciton eigenstates is solely determined by the strength of the tunnel coupling $t$. Since we work in a regime where the electron can hybridize while the hole remains confined in the upper QD\,\cite{Bracker2006}, the energy splitting between the BO and AB eigenstates $\Delta E$ corresponds to the energy difference between the observed neutral exciton eigenstates. This allows us to analyze the magnetic field-dependent energy separation between the BO and AB wave functions and thus obtain the tunnel coupling between the QDs.

\begin{figure}
\includegraphics{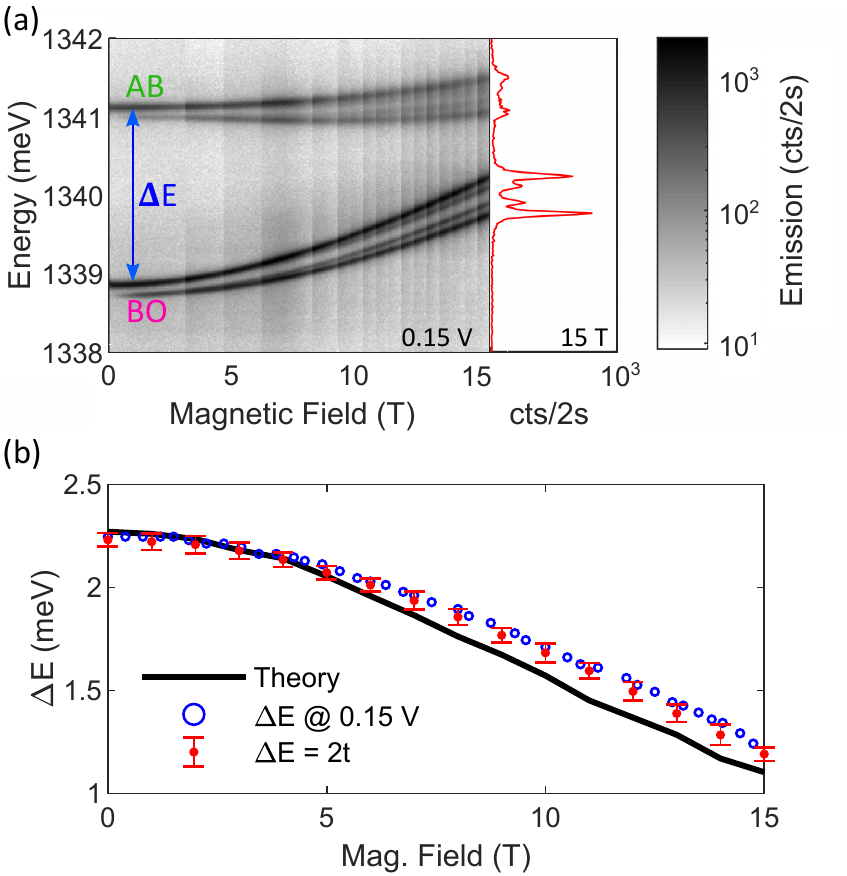}
\caption{\label{fig:2} (a) Magneto-photoluminescence measurement in the tunnel-coupling regime at 0.15\,V. The energy splitting $\Delta E$ between the bonding (BO) and anti-bonding (AB) eigenstate is illustrated by a blue arrow. The red solid line shows the emission spectrum at 15\,T. (b) $\Delta E$, the energy splitting, as function of the magnetic field. The blue data points indicate the energy splitting between the highest energy transitions of the BO and the AB eigenstate at 0.15\,V. The red data points are extracted from fitting the voltage-dependent energy splitting. The black solid line shows the numerically predicted energy splitting $\Delta E$.}
\end{figure}

To analyze the in-situ tunability of the tunnel coupling, we apply an in-plane magnetic field along the x-direction (Voigt geometry). Figure\,\ref{fig:2}\,(a) shows a magneto-photoluminescence measurement of the BO and AB state at the lowest splitting of 0.15\,V. In addition to a diamagnetic shift\,\cite{Bayer1998}, the magnetic field leads to a mixing of bright and dark exciton states \cite{Bayer2000} and a Zeeman splitting of the spin states \cite{Bayer2002}. This results in the occurrence of the four allowed optical transitions for both the BO and AB eigenstates, as shown by the single photoluminescence trace in Figure \ref{fig:2}\,(a) recorded at 15T.  . 
With an increasing magnetic field, the energy separation $\Delta E$ between the eigenstates with the same spin configuration reduces. We define $\Delta E$ as the difference between the highest energy transition of the BO and of the corresponding AB eigenstate.

Figure\,\ref{fig:2}\,(b) shows the energy splitting $\Delta E$ as a function of magnetic field. The blue data points describe the energy separation at 0.15\,V obtained from Figure\,\ref{fig:2}\,(a). The energy separation between the two eigenstates decreases with increasing magnetic field.
In order to unambiguously identify the observed transitions, we complement the experimental results with numerical calculations, performed within the envelope-function and the effective-mass approach \cite{jacak2013quantum}. The single particle energies and eigenstates are obtained by diagonalizing on a real-space grid of $524\,288$ points
a Hamiltonian that includes a kinetic and a potential contribution:
\begin{equation}
\label{equ:Hamiltonian3D}
H_\chi(\textbf{r}) = \frac{1}{2m_\chi^*}\left(-i\hbar\nabla+\frac{q_\chi}{c}\textbf{A} \right)^2+V_\chi(\textbf{r}),
\end{equation}
where $\textbf{A}=\frac{1}{2}\textbf{B}\times\textbf{r}$ is the vector potential, $\chi=e,h$ specifies the particle type and its charge ($q_h=-q_e=|e|$), and $m_\chi^*$ is the effective mass. The confinement potential includes a parabolic in-plane contribution, 
$V_{xy,\chi}(x,y)=\frac{1}{2}m_\chi^*\omega_\chi^2(x^2+y^2)$,
and a double square well in the vertical ($z$) direction. In particular, the two wells have widths of $d_1=2.7$\,nm and $d_2=2.9$\,nm, matching the height of the bottom and top QD, respectively. We note that the dot heights were precisely fixed at these values using the In-flush method\,\cite{Wasilewski1999}.  Similarly, the barrier between the two dots was defined during growth and fixed at $l=7.3$\,nm. The height of the barrier is set to $V_0=690$\,meV, corresponding to the conduction band offset of strained InAs and GaAs\,\cite{Colombelli2000}. In addition, the potential includes a linear term $V_{bias,\chi} = q_\chi \mathcal{E} z$, with $\mathcal{E}=3.63$\,meV/nm to account for the electric field. 

We include the Coulomb interaction perturbatively, and identify the energy of the lowest (first excited) exciton state with the sum of the electron ground state energy, the hole ground state energy, and the direct Coulomb matrix element involving these single-particle states. Further details are provided in the Supplemental Material\,\cite{Supp}.

The black solid curve in Figure\,\ref{fig:2}\,(b) shows the numerically obtained energy splitting $\Delta E$ as a function of the strength of the in-plane magnetic field. We obtain an excellent qualitative agreement between experimental and theoretical results. Quantitative differences may result from simplifications in the applied three-dimensional potential.

The magnetic field leads to an enhanced confinement of the charges in the potential wells along the growth direction $z$, which is associated with a reduction of the electron wave function density inside the tunneling barrier. Thus, the coupling strength decreases with increasing magnetic field. In the presented magnetic field range we observe a decrease of the splitting of the avoided crossing by ($53.4\pm 1.7$)\,\%, which corresponds to an absolute energy shift of $(-1.04\pm0.05)$\,meV and a change of the tunnel coupling strength from $t = (1.12\pm0.02)$\,meV to $(0.60\pm0.02)$\,meV. This is a factor of 21$\times$ higher than what could be previously achieved by strain tuning the device\,\cite{Zallo2014}.

\begin{figure}
\includegraphics{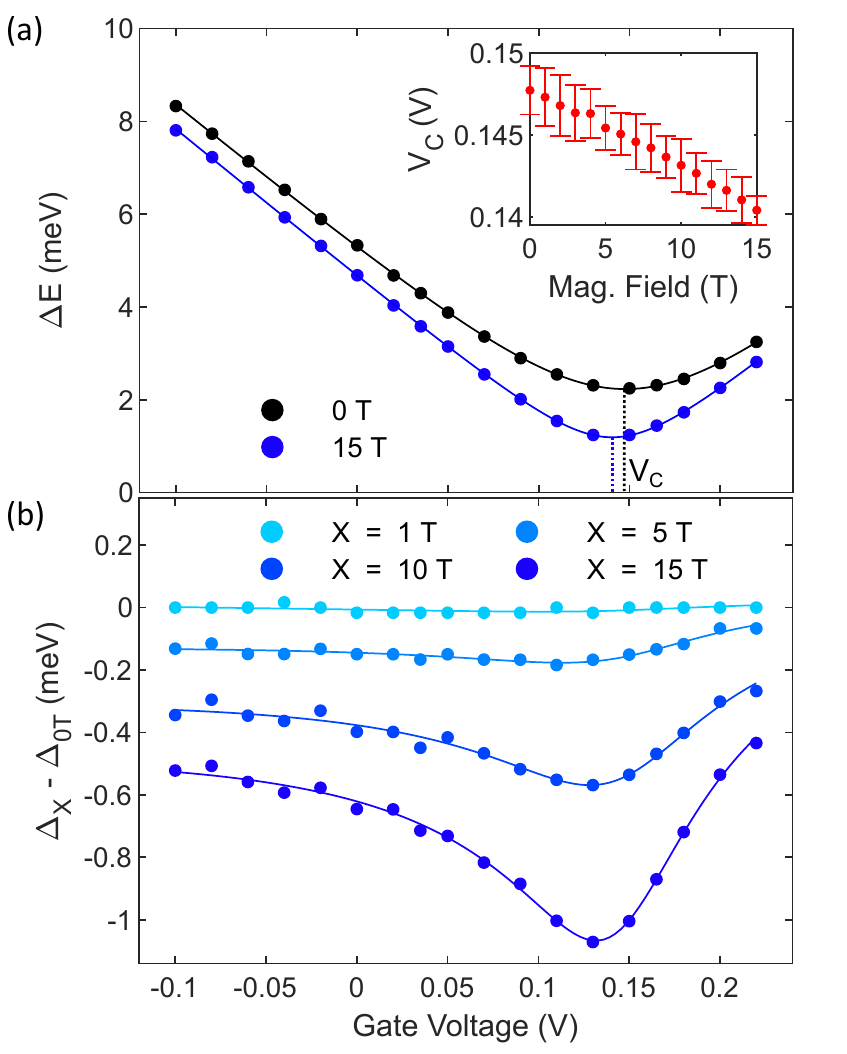}
\caption{\label{fig:3} (a) $\Delta E$ as a function of the gate voltage at 0\,T (black dots) and 15\,T (blue dots). Solid lines visualize the fit of the energy splitting. $V_C$ describes the voltage, where $\Delta E$ is minimal. The inset shows $V_C$ as function of the magnetic field. (b) Difference of $\Delta E$ with and without applied magnetic field as a function of the gate voltage. The solid lines show the difference between the fits applied in (a).}
\end{figure}

As the magnetic field is increased, the decrease in $\Delta E$ and hence the tunnel coupling should be maximal in the center of the avoided crossing. Thus, we continue to analyze $\Delta E$ close to and away from the avoided crossing to show conclusively that the reduction of $\Delta E$ is indeed caused by a reduction of the tunnel coupling. Therefore, we measured $\Delta E$ for different DC gate voltages tracking over the avoided crossing.

Figure\,\ref{fig:3}\,(a) shows $\Delta E$ as a function of the gate voltage around the avoided crossing at 0\,T (black points) and 15\,T (blue points). The solid lines result from fitting the data with the difference of the eigenenergies calculated from the coupled two-state model. Both fits show a minimum at the center voltage of the avoided crossing, depicted by the vertical dashed lines. $V_C$ marks the center voltage of the avoided crossing at 0\,T. As before, $\Delta E$ is reduced at 15\,T compared to 0\,T. This results in a reduced tunneling coupling strength, represented by the fit parameter $t$.
At $V_C$ the energy splitting is $\Delta E=2t$, which allows the extraction of the minimum splitting from the applied fit. The red data points in Figure\,\ref{fig:2}\,(b) show the minimum energy splitting for different magnetic fields. While the fitted and measured $\Delta E$ are in excellent agreement at low magnetic fields, a small deviation is observed for higher fields. This is caused by a magnetic field-dependent shift of $V_C$. The inset in Figure\,\ref{fig:3}\,(a) shows the evolution of $V_C$ with increasing magnetic field. The observed voltage shift of the tunnel coupling originates from the different sizes of the two individual QDs. By increasing the magnetic field and thus the confinement of the charge, the eigenenergies of the two QDs encounter different energy shifts. As a result, a lower gate voltage is required to attain the coupling condition. Thus, $V_C$ decreases with increasing magnetic field.

To demonstrate the reduction in energy splitting at the avoided crossing, we analyze the difference of $\Delta E$ with and without an applied magnetic field. 
Figure\,\ref{fig:3}\,(b) shows the voltage-dependent difference $\Delta_{X\,\text{T}}$ - $\Delta_{0\,\text{T}}$, with $X\in\{1,5,10,15\}$. The fitted curves (solid lines) are obtained by subtracting the difference of the fitted eigenenergies in Figure\,\ref{fig:3}\,(a). The asymmetry of the curve is caused by the change of $V_C$ with increasing magnetic field, as discussed before.
The data presented exhibits a minimum of the energy difference at the avoided crossing. This minimum confirms that the variation of the energy splitting is indeed maximal at the avoided crossing. Thus, we can conclude that the applied magnetic field reduces the tunnel coupling between the two quantum dots.
\\

In summary, we have demonstrated a reduction in the energy splitting between BO and AB eigenstates of a QDM by applying in-plane magnetic fields. This corresponds to a reduction in the tunnel coupling strength between the two QDs. Fitting the energy difference between the eigenstates of the system to the voltage and magnetic field-dependent measurements allows quantification of the variation of the tunnel coupling strength. We obtain a wide tuning range of $(53.4\pm 1.7)$\,\% with a maximum tuning rate of $(-101 \pm 6)$\,$\mu$eV/T. The decrease of the absolute energy splitting is $(1.04\pm0.05)$\,meV. Our results confirm several theoretical predictions \cite{Burkard2000,Korkusinski2001,Bellucci2004,Jacob2004,DiasDaSilva2007,Climente2008} and demonstrate that in-plane magnetic fields enable the tuning of the tunnel coupling in QDMs.

The tunability of the tunnel coupling is crucial for controlling the eigenstates of solid-state qubits in QDMs. The presented results allow optimization of the gate operations for the generation of two-dimensional photonic cluster states\,\cite{Economou2010}. Moreover, tuning the eigenenergies of a qubit facilitates its synchronization with other qubits and thus their entanglement to enable multi-qubit networks. These results are generally applicable to other artificial molecules, e.g. in two-dimensional materials with a similar trapping potential\,\cite{Burkard2000}.

\begin{acknowledgments}
The authors gratefully acknowledge financial support from the German Federal Ministry of Education and Research (BMBF) via Q.Link.X (16KIS0874, 16KIS086), QR.X (16KISQ027, 16KISQ014, 16KISQ012 and 16KISQ009), MOQUA (13N14846), the Horizon 2020 research and innovation program of the European Union under grant agreement 862035 (QLUSTER) and the Deutsche Forschungsgemeinschaft (DFG, German Research Foundation) via SQAM (FI947-5-1), DIP (FI947-6-1), and the Excellence Cluster MCQST (EXC-2111, 390814868). F.B. gratefully acknowledges the Exploring Quantum Matter (ExQM) programme funded by the State of Bavaria. FT and EM acknowledge financial support from the European Commission through the project IQubits (Call: H2020–FETOPEN–2018–2019–2020–01, ProjectID: 829005).
\end{acknowledgments}


\bibliography{main.bbl}

\begin{thebibliography}{34}%
\makeatletter
\providecommand \@ifxundefined [1]{%
 \@ifx{#1\undefined}
}%
\providecommand \@ifnum [1]{%
 \ifnum #1\expandafter \@firstoftwo
 \else \expandafter \@secondoftwo
 \fi
}%
\providecommand \@ifx [1]{%
 \ifx #1\expandafter \@firstoftwo
 \else \expandafter \@secondoftwo
 \fi
}%
\providecommand \natexlab [1]{#1}%
\providecommand \enquote  [1]{``#1''}%
\providecommand \bibnamefont  [1]{#1}%
\providecommand \bibfnamefont [1]{#1}%
\providecommand \citenamefont [1]{#1}%
\providecommand \href@noop [0]{\@secondoftwo}%
\providecommand \href [0]{\begingroup \@sanitize@url \@href}%
\providecommand \@href[1]{\@@startlink{#1}\@@href}%
\providecommand \@@href[1]{\endgroup#1\@@endlink}%
\providecommand \@sanitize@url [0]{\catcode `\\12\catcode `\$12\catcode
  `\&12\catcode `\#12\catcode `\^12\catcode `\_12\catcode `\%12\relax}%
\providecommand \@@startlink[1]{}%
\providecommand \@@endlink[0]{}%
\providecommand \url  [0]{\begingroup\@sanitize@url \@url }%
\providecommand \@url [1]{\endgroup\@href {#1}{\urlprefix }}%
\providecommand \urlprefix  [0]{URL }%
\providecommand \Eprint [0]{\href }%
\providecommand \doibase [0]{https://doi.org/}%
\providecommand \selectlanguage [0]{\@gobble}%
\providecommand \bibinfo  [0]{\@secondoftwo}%
\providecommand \bibfield  [0]{\@secondoftwo}%
\providecommand \translation [1]{[#1]}%
\providecommand \BibitemOpen [0]{}%
\providecommand \bibitemStop [0]{}%
\providecommand \bibitemNoStop [0]{.\EOS\space}%
\providecommand \EOS [0]{\spacefactor3000\relax}%
\providecommand \BibitemShut  [1]{\csname bibitem#1\endcsname}%
\let\auto@bib@innerbib\@empty
\bibitem [{\citenamefont {Lidar}\ \emph {et~al.}(1998)\citenamefont {Lidar},
  \citenamefont {Chuang},\ and\ \citenamefont {Whaley}}]{Lidar1998}%
  \BibitemOpen
  \bibfield  {author} {\bibinfo {author} {\bibfnamefont {D.~A.}\ \bibnamefont
  {Lidar}}, \bibinfo {author} {\bibfnamefont {I.~L.}\ \bibnamefont {Chuang}},\
  and\ \bibinfo {author} {\bibfnamefont {K.~B.}\ \bibnamefont {Whaley}},\
  }\bibfield  {title} {\bibinfo {title} {Decoherence-free subspaces for quantum
  computation},\ }\href@noop {} {\bibfield  {journal} {\bibinfo  {journal}
  {Physical Review Letters}\ }\textbf {\bibinfo {volume} {81}},\ \bibinfo
  {pages} {2594} (\bibinfo {year} {1998})}\BibitemShut {NoStop}%
\bibitem [{\citenamefont {Douçot}\ and\ \citenamefont
  {Ioffe}(2012)}]{Doucot2012}%
  \BibitemOpen
  \bibfield  {author} {\bibinfo {author} {\bibfnamefont {B.}~\bibnamefont
  {Douçot}}\ and\ \bibinfo {author} {\bibfnamefont {L.~B.}\ \bibnamefont
  {Ioffe}},\ }\bibfield  {title} {\bibinfo {title} {Physical implementation of
  protected qubits},\ }\href
  {https://iopscience.iop.org/article/10.1088/0034-4885/75/7/072001
  https://iopscience.iop.org/article/10.1088/0034-4885/75/7/072001/meta}
  {\bibfield  {journal} {\bibinfo  {journal} {Reports on Progress in Physics}\
  }\textbf {\bibinfo {volume} {75}},\ \bibinfo {pages} {072001} (\bibinfo
  {year} {2012})}\BibitemShut {NoStop}%
\bibitem [{\citenamefont {Petta}\ \emph {et~al.}(2005)\citenamefont {Petta},
  \citenamefont {Johnson}, \citenamefont {Taylor}, \citenamefont {Laird},
  \citenamefont {Yacoby}, \citenamefont {Lukin}, \citenamefont {Marcus},
  \citenamefont {Hanson},\ and\ \citenamefont {Gossard}}]{Petta2005}%
  \BibitemOpen
  \bibfield  {author} {\bibinfo {author} {\bibfnamefont {J.~R.}\ \bibnamefont
  {Petta}}, \bibinfo {author} {\bibfnamefont {A.~C.}\ \bibnamefont {Johnson}},
  \bibinfo {author} {\bibfnamefont {J.~M.}\ \bibnamefont {Taylor}}, \bibinfo
  {author} {\bibfnamefont {E.~A.}\ \bibnamefont {Laird}}, \bibinfo {author}
  {\bibfnamefont {A.}~\bibnamefont {Yacoby}}, \bibinfo {author} {\bibfnamefont
  {M.~D.}\ \bibnamefont {Lukin}}, \bibinfo {author} {\bibfnamefont {C.~M.}\
  \bibnamefont {Marcus}}, \bibinfo {author} {\bibfnamefont {M.~P.}\
  \bibnamefont {Hanson}},\ and\ \bibinfo {author} {\bibfnamefont {A.~C.}\
  \bibnamefont {Gossard}},\ }\bibfield  {title} {\bibinfo {title} {Coherent
  manipulation of coupled electron spins in semiconductor quantum dots},\
  }\href {http://science.sciencemag.org/} {\bibfield  {journal} {\bibinfo
  {journal} {Science}\ }\textbf {\bibinfo {volume} {309}},\ \bibinfo {pages}
  {2180} (\bibinfo {year} {2005})}\BibitemShut {NoStop}%
\bibitem [{\citenamefont {Stopa}\ and\ \citenamefont
  {Marcus}(2008)}]{Stopa2008}%
  \BibitemOpen
  \bibfield  {author} {\bibinfo {author} {\bibfnamefont {M.}~\bibnamefont
  {Stopa}}\ and\ \bibinfo {author} {\bibfnamefont {C.~M.}\ \bibnamefont
  {Marcus}},\ }\bibfield  {title} {\bibinfo {title} {Magnetic field control of
  exchange and noise immunity in double quantum dots},\ }\href@noop {}
  {\bibfield  {journal} {\bibinfo  {journal} {Nano Letters}\ }\textbf {\bibinfo
  {volume} {8}},\ \bibinfo {pages} {1778} (\bibinfo {year} {2008})}\BibitemShut
  {NoStop}%
\bibitem [{\citenamefont {Weiss}\ \emph {et~al.}(2012)\citenamefont {Weiss},
  \citenamefont {Elzerman}, \citenamefont {Delley}, \citenamefont
  {Miguel-Sanchez},\ and\ \citenamefont {Imamo\ifmmode~\breve{g}\else
  \u{g}\fi{}lu}}]{Weiss2012}%
  \BibitemOpen
  \bibfield  {author} {\bibinfo {author} {\bibfnamefont {K.~M.}\ \bibnamefont
  {Weiss}}, \bibinfo {author} {\bibfnamefont {J.~M.}\ \bibnamefont {Elzerman}},
  \bibinfo {author} {\bibfnamefont {Y.~L.}\ \bibnamefont {Delley}}, \bibinfo
  {author} {\bibfnamefont {J.}~\bibnamefont {Miguel-Sanchez}},\ and\ \bibinfo
  {author} {\bibfnamefont {A.}~\bibnamefont {Imamo\ifmmode~\breve{g}\else
  \u{g}\fi{}lu}},\ }\bibfield  {title} {\bibinfo {title} {Coherent two-electron
  spin qubits in an optically active pair of coupled ingaas quantum dots},\
  }\href {https://doi.org/10.1103/PhysRevLett.109.107401} {\bibfield  {journal}
  {\bibinfo  {journal} {Phys. Rev. Lett.}\ }\textbf {\bibinfo {volume} {109}},\
  \bibinfo {pages} {107401} (\bibinfo {year} {2012})}\BibitemShut {NoStop}%
\bibitem [{\citenamefont {Tran}\ \emph {et~al.}(2022)\citenamefont {Tran},
  \citenamefont {Bracker}, \citenamefont {Yakes}, \citenamefont {Grim},\ and\
  \citenamefont {Carter}}]{Tran2022}%
  \BibitemOpen
  \bibfield  {author} {\bibinfo {author} {\bibfnamefont {K.~X.}\ \bibnamefont
  {Tran}}, \bibinfo {author} {\bibfnamefont {A.~S.}\ \bibnamefont {Bracker}},
  \bibinfo {author} {\bibfnamefont {M.~K.}\ \bibnamefont {Yakes}}, \bibinfo
  {author} {\bibfnamefont {J.~Q.}\ \bibnamefont {Grim}},\ and\ \bibinfo
  {author} {\bibfnamefont {S.~G.}\ \bibnamefont {Carter}},\ }\bibfield  {title}
  {\bibinfo {title} {Enhanced spin coherence of a self-assembled quantum dot
  molecule at the optimal electrical bias},\ }\href@noop {} {\bibfield
  {journal} {\bibinfo  {journal} {Phys. Rev. Lett.}\ }\textbf {\bibinfo
  {volume} {129}},\ \bibinfo {pages} {027403} (\bibinfo {year}
  {2022})}\BibitemShut {NoStop}%
\bibitem [{\citenamefont {Benny}\ \emph {et~al.}(2011)\citenamefont {Benny},
  \citenamefont {Khatsevich}, \citenamefont {Kodriano}, \citenamefont {Poem},
  \citenamefont {Presman}, \citenamefont {Galushko}, \citenamefont {Petroff},\
  and\ \citenamefont {Gershoni}}]{Benny2011}%
  \BibitemOpen
  \bibfield  {author} {\bibinfo {author} {\bibfnamefont {Y.}~\bibnamefont
  {Benny}}, \bibinfo {author} {\bibfnamefont {S.}~\bibnamefont {Khatsevich}},
  \bibinfo {author} {\bibfnamefont {Y.}~\bibnamefont {Kodriano}}, \bibinfo
  {author} {\bibfnamefont {E.}~\bibnamefont {Poem}}, \bibinfo {author}
  {\bibfnamefont {R.}~\bibnamefont {Presman}}, \bibinfo {author} {\bibfnamefont
  {D.}~\bibnamefont {Galushko}}, \bibinfo {author} {\bibfnamefont {P.~M.}\
  \bibnamefont {Petroff}},\ and\ \bibinfo {author} {\bibfnamefont
  {D.}~\bibnamefont {Gershoni}},\ }\bibfield  {title} {\bibinfo {title}
  {Coherent optical writing and reading of the exciton spin state in single
  quantum dots},\ }\href
  {https://journals.aps.org/prl/abstract/10.1103/PhysRevLett.106.040504}
  {\bibfield  {journal} {\bibinfo  {journal} {Physical Review Letters}\
  }\textbf {\bibinfo {volume} {106}},\ \bibinfo {pages} {040504} (\bibinfo
  {year} {2011})}\BibitemShut {NoStop}%
\bibitem [{\citenamefont {Favero}\ \emph {et~al.}(2003)\citenamefont {Favero},
  \citenamefont {Cassabois}, \citenamefont {Ferreira}, \citenamefont {Darson},
  \citenamefont {Voisin}, \citenamefont {Tignon}, \citenamefont {Delalande},
  \citenamefont {Bastard}, \citenamefont {Roussignol},\ and\ \citenamefont
  {Gérard}}]{Favero2003}%
  \BibitemOpen
  \bibfield  {author} {\bibinfo {author} {\bibfnamefont {I.}~\bibnamefont
  {Favero}}, \bibinfo {author} {\bibfnamefont {G.}~\bibnamefont {Cassabois}},
  \bibinfo {author} {\bibfnamefont {R.}~\bibnamefont {Ferreira}}, \bibinfo
  {author} {\bibfnamefont {D.}~\bibnamefont {Darson}}, \bibinfo {author}
  {\bibfnamefont {C.}~\bibnamefont {Voisin}}, \bibinfo {author} {\bibfnamefont
  {J.}~\bibnamefont {Tignon}}, \bibinfo {author} {\bibfnamefont
  {C.}~\bibnamefont {Delalande}}, \bibinfo {author} {\bibfnamefont
  {G.}~\bibnamefont {Bastard}}, \bibinfo {author} {\bibfnamefont
  {P.}~\bibnamefont {Roussignol}},\ and\ \bibinfo {author} {\bibfnamefont
  {J.~M.}\ \bibnamefont {Gérard}},\ }\bibfield  {title} {\bibinfo {title}
  {Acoustic phonon sidebands in the emission line of single {InAs/GaAs} quantum
  dots},\ }\href
  {https://journals.aps.org/prb/abstract/10.1103/PhysRevB.68.233301} {\bibfield
   {journal} {\bibinfo  {journal} {Physical Review B - Condensed Matter and
  Materials Physics}\ }\textbf {\bibinfo {volume} {68}},\ \bibinfo {pages}
  {233301} (\bibinfo {year} {2003})}\BibitemShut {NoStop}%
\bibitem [{\citenamefont {Bracker}\ \emph {et~al.}(2006)\citenamefont
  {Bracker}, \citenamefont {Scheibner}, \citenamefont {Doty}, \citenamefont
  {Stinaff}, \citenamefont {Ponomarev}, \citenamefont {Kim}, \citenamefont
  {Whitman}, \citenamefont {Reinecke},\ and\ \citenamefont
  {Gammon}}]{Bracker2006}%
  \BibitemOpen
  \bibfield  {author} {\bibinfo {author} {\bibfnamefont {A.~S.}\ \bibnamefont
  {Bracker}}, \bibinfo {author} {\bibfnamefont {M.}~\bibnamefont {Scheibner}},
  \bibinfo {author} {\bibfnamefont {M.~F.}\ \bibnamefont {Doty}}, \bibinfo
  {author} {\bibfnamefont {E.~A.}\ \bibnamefont {Stinaff}}, \bibinfo {author}
  {\bibfnamefont {I.~V.}\ \bibnamefont {Ponomarev}}, \bibinfo {author}
  {\bibfnamefont {J.~C.}\ \bibnamefont {Kim}}, \bibinfo {author} {\bibfnamefont
  {L.~J.}\ \bibnamefont {Whitman}}, \bibinfo {author} {\bibfnamefont {T.~L.}\
  \bibnamefont {Reinecke}},\ and\ \bibinfo {author} {\bibfnamefont
  {D.}~\bibnamefont {Gammon}},\ }\bibfield  {title} {\bibinfo {title}
  {Engineering electron and hole tunneling with asymmetric {InAs} quantum dot
  molecules},\ }\href {http://aip.scitation.org/doi/10.1063/1.2400397}
  {\bibfield  {journal} {\bibinfo  {journal} {Applied Physics Letters}\
  }\textbf {\bibinfo {volume} {89}},\ \bibinfo {pages} {233110} (\bibinfo
  {year} {2006})}\BibitemShut {NoStop}%
\bibitem [{\citenamefont {Martins}\ \emph {et~al.}(2016)\citenamefont
  {Martins}, \citenamefont {Malinowski}, \citenamefont {Nissen}, \citenamefont
  {Barnes}, \citenamefont {Fallahi}, \citenamefont {Gardner}, \citenamefont
  {Manfra}, \citenamefont {Marcus},\ and\ \citenamefont
  {Kuemmeth}}]{Martins2016}%
  \BibitemOpen
  \bibfield  {author} {\bibinfo {author} {\bibfnamefont {F.}~\bibnamefont
  {Martins}}, \bibinfo {author} {\bibfnamefont {F.~K.}\ \bibnamefont
  {Malinowski}}, \bibinfo {author} {\bibfnamefont {P.~D.}\ \bibnamefont
  {Nissen}}, \bibinfo {author} {\bibfnamefont {E.}~\bibnamefont {Barnes}},
  \bibinfo {author} {\bibfnamefont {S.}~\bibnamefont {Fallahi}}, \bibinfo
  {author} {\bibfnamefont {G.~C.}\ \bibnamefont {Gardner}}, \bibinfo {author}
  {\bibfnamefont {M.~J.}\ \bibnamefont {Manfra}}, \bibinfo {author}
  {\bibfnamefont {C.~M.}\ \bibnamefont {Marcus}},\ and\ \bibinfo {author}
  {\bibfnamefont {F.}~\bibnamefont {Kuemmeth}},\ }\bibfield  {title} {\bibinfo
  {title} {Noise suppression using symmetric exchange gates in spin qubits},\
  }\href {https://journals.aps.org/prl/abstract/10.1103/PhysRevLett.116.116801}
  {\bibfield  {journal} {\bibinfo  {journal} {Physical Review Letters}\
  }\textbf {\bibinfo {volume} {116}},\ \bibinfo {pages} {116801} (\bibinfo
  {year} {2016})}\BibitemShut {NoStop}%
\bibitem [{\citenamefont {Kim}\ \emph {et~al.}(2011)\citenamefont {Kim},
  \citenamefont {Carter}, \citenamefont {Greilich}, \citenamefont {Bracker},\
  and\ \citenamefont {Gammon}}]{Kim2011}%
  \BibitemOpen
  \bibfield  {author} {\bibinfo {author} {\bibfnamefont {D.}~\bibnamefont
  {Kim}}, \bibinfo {author} {\bibfnamefont {S.~G.}\ \bibnamefont {Carter}},
  \bibinfo {author} {\bibfnamefont {A.}~\bibnamefont {Greilich}}, \bibinfo
  {author} {\bibfnamefont {A.~S.}\ \bibnamefont {Bracker}},\ and\ \bibinfo
  {author} {\bibfnamefont {D.}~\bibnamefont {Gammon}},\ }\bibfield  {title}
  {\bibinfo {title} {Ultrafast optical control of entanglement between two
  quantum-dot spins},\ }\href {http://www.nature.com/articles/nphys1863}
  {\bibfield  {journal} {\bibinfo  {journal} {Nature Physics}\ }\textbf
  {\bibinfo {volume} {7}},\ \bibinfo {pages} {223} (\bibinfo {year}
  {2011})}\BibitemShut {NoStop}%
\bibitem [{\citenamefont {Kimble}(2008)}]{Kimble2008}%
  \BibitemOpen
  \bibfield  {author} {\bibinfo {author} {\bibfnamefont {H.~J.}\ \bibnamefont
  {Kimble}},\ }\bibfield  {title} {\bibinfo {title} {The quantum internet},\
  }\href@noop {} {\bibfield  {journal} {\bibinfo  {journal} {Nature}\ }\textbf
  {\bibinfo {volume} {453}},\ \bibinfo {pages} {1023} (\bibinfo {year}
  {2008})}\BibitemShut {NoStop}%
\bibitem [{\citenamefont {Economou}\ \emph {et~al.}(2010)\citenamefont
  {Economou}, \citenamefont {Lindner},\ and\ \citenamefont
  {Rudolph}}]{Economou2010}%
  \BibitemOpen
  \bibfield  {author} {\bibinfo {author} {\bibfnamefont {S.~E.}\ \bibnamefont
  {Economou}}, \bibinfo {author} {\bibfnamefont {N.}~\bibnamefont {Lindner}},\
  and\ \bibinfo {author} {\bibfnamefont {T.}~\bibnamefont {Rudolph}},\
  }\bibfield  {title} {\bibinfo {title} {Optically generated 2-dimensional
  photonic cluster state from coupled quantum dots},\ }\href@noop {} {\bibfield
   {journal} {\bibinfo  {journal} {Phys. Rev. Lett.}\ }\textbf {\bibinfo
  {volume} {105}},\ \bibinfo {pages} {093601} (\bibinfo {year}
  {2010})}\BibitemShut {NoStop}%
\bibitem [{\citenamefont {Raussendorf}\ and\ \citenamefont
  {Briegel}(2001)}]{Raussendorf2001}%
  \BibitemOpen
  \bibfield  {author} {\bibinfo {author} {\bibfnamefont {R.}~\bibnamefont
  {Raussendorf}}\ and\ \bibinfo {author} {\bibfnamefont {H.~J.}\ \bibnamefont
  {Briegel}},\ }\bibfield  {title} {\bibinfo {title} {A one-way quantum
  computer},\ }\href@noop {} {\bibfield  {journal} {\bibinfo  {journal}
  {Physical Review Letters}\ }\textbf {\bibinfo {volume} {86}},\ \bibinfo
  {pages} {5188} (\bibinfo {year} {2001})}\BibitemShut {NoStop}%
\bibitem [{\citenamefont {Zallo}\ \emph {et~al.}(2014)\citenamefont {Zallo},
  \citenamefont {Trotta}, \citenamefont {Křápek}, \citenamefont {Huo},
  \citenamefont {Atkinson}, \citenamefont {Ding}, \citenamefont {Sikola},
  \citenamefont {Rastelli},\ and\ \citenamefont {Schmidt}}]{Zallo2014}%
  \BibitemOpen
  \bibfield  {author} {\bibinfo {author} {\bibfnamefont {E.}~\bibnamefont
  {Zallo}}, \bibinfo {author} {\bibfnamefont {R.}~\bibnamefont {Trotta}},
  \bibinfo {author} {\bibfnamefont {V.}~\bibnamefont {Křápek}}, \bibinfo
  {author} {\bibfnamefont {Y.~H.}\ \bibnamefont {Huo}}, \bibinfo {author}
  {\bibfnamefont {P.}~\bibnamefont {Atkinson}}, \bibinfo {author}
  {\bibfnamefont {F.}~\bibnamefont {Ding}}, \bibinfo {author} {\bibfnamefont
  {T.~Ë.}\ \bibnamefont {Sikola}}, \bibinfo {author} {\bibfnamefont
  {A.}~\bibnamefont {Rastelli}},\ and\ \bibinfo {author} {\bibfnamefont
  {O.~G.}\ \bibnamefont {Schmidt}},\ }\bibfield  {title} {\bibinfo {title}
  {Strain-induced active tuning of the coherent tunneling in quantum dot
  molecules},\ }\href@noop {} {\bibfield  {journal} {\bibinfo  {journal}
  {Physical Review B}\ }\textbf {\bibinfo {volume} {89}},\ \bibinfo {pages}
  {241303} (\bibinfo {year} {2014})}\BibitemShut {NoStop}%
\bibitem [{\citenamefont {Carter}\ \emph {et~al.}(2019)\citenamefont {Carter},
  \citenamefont {Bracker}, \citenamefont {Yakes}, \citenamefont {Zalalutdinov},
  \citenamefont {Kim}, \citenamefont {Kim}, \citenamefont {Lee},\ and\
  \citenamefont {Gammon}}]{Carter2019}%
  \BibitemOpen
  \bibfield  {author} {\bibinfo {author} {\bibfnamefont {S.~G.}\ \bibnamefont
  {Carter}}, \bibinfo {author} {\bibfnamefont {A.~S.}\ \bibnamefont {Bracker}},
  \bibinfo {author} {\bibfnamefont {M.~K.}\ \bibnamefont {Yakes}}, \bibinfo
  {author} {\bibfnamefont {M.~K.}\ \bibnamefont {Zalalutdinov}}, \bibinfo
  {author} {\bibfnamefont {M.}~\bibnamefont {Kim}}, \bibinfo {author}
  {\bibfnamefont {C.~S.}\ \bibnamefont {Kim}}, \bibinfo {author} {\bibfnamefont
  {B.}~\bibnamefont {Lee}},\ and\ \bibinfo {author} {\bibfnamefont
  {D.}~\bibnamefont {Gammon}},\ }\bibfield  {title} {\bibinfo {title} {Tunable
  coupling of a double quantum dot spin system to a mechanical resonator},\
  }\href {https://pubs.acs.org/sharingguidelines} {\bibfield  {journal}
  {\bibinfo  {journal} {Nano Lett}\ }\textbf {\bibinfo {volume} {19}},\
  \bibinfo {pages} {11} (\bibinfo {year} {2019})}\BibitemShut {NoStop}%
\bibitem [{\citenamefont {Burkard}\ \emph {et~al.}(2000)\citenamefont
  {Burkard}, \citenamefont {Seelig},\ and\ \citenamefont {Loss}}]{Burkard2000}%
  \BibitemOpen
  \bibfield  {author} {\bibinfo {author} {\bibfnamefont {G.}~\bibnamefont
  {Burkard}}, \bibinfo {author} {\bibfnamefont {G.}~\bibnamefont {Seelig}},\
  and\ \bibinfo {author} {\bibfnamefont {D.}~\bibnamefont {Loss}},\ }\bibfield
  {title} {\bibinfo {title} {Spin interactions and switching in vertically
  tunnel-coupled quantum dots},\ }\href@noop {} {\bibfield  {journal} {\bibinfo
   {journal} {Phys. Rev. B}\ }\textbf {\bibinfo {volume} {62}},\ \bibinfo
  {pages} {2581} (\bibinfo {year} {2000})}\BibitemShut {NoStop}%
\bibitem [{\citenamefont {Korkusi\ifmmode~\acute{n}\else \'{n}\fi{}ski}\ and\
  \citenamefont {Hawrylak}(2001)}]{Korkusinski2001}%
  \BibitemOpen
  \bibfield  {author} {\bibinfo {author} {\bibfnamefont {M.}~\bibnamefont
  {Korkusi\ifmmode~\acute{n}\else \'{n}\fi{}ski}}\ and\ \bibinfo {author}
  {\bibfnamefont {P.}~\bibnamefont {Hawrylak}},\ }\bibfield  {title} {\bibinfo
  {title} {Electronic structure of vertically stacked self-assembled quantum
  disks},\ }\href@noop {} {\bibfield  {journal} {\bibinfo  {journal} {Phys.
  Rev. B}\ }\textbf {\bibinfo {volume} {63}},\ \bibinfo {pages} {195311}
  (\bibinfo {year} {2001})}\BibitemShut {NoStop}%
\bibitem [{\citenamefont {Bellucci}\ \emph {et~al.}(2004)\citenamefont
  {Bellucci}, \citenamefont {Troiani}, \citenamefont {Goldoni},\ and\
  \citenamefont {Molinari}}]{Bellucci2004}%
  \BibitemOpen
  \bibfield  {author} {\bibinfo {author} {\bibfnamefont {D.}~\bibnamefont
  {Bellucci}}, \bibinfo {author} {\bibfnamefont {F.}~\bibnamefont {Troiani}},
  \bibinfo {author} {\bibfnamefont {G.}~\bibnamefont {Goldoni}},\ and\ \bibinfo
  {author} {\bibfnamefont {E.}~\bibnamefont {Molinari}},\ }\bibfield  {title}
  {\bibinfo {title} {Neutral and charged electron-hole complexes in artificial
  molecules: Quantum transitions induced by the in-plane magnetic field},\
  }\href@noop {} {\bibfield  {journal} {\bibinfo  {journal} {Phys. Rev. B}\
  }\textbf {\bibinfo {volume} {70}},\ \bibinfo {pages} {205332} (\bibinfo
  {year} {2004})}\BibitemShut {NoStop}%
\bibitem [{\citenamefont {Jacob}\ \emph {et~al.}(2004)\citenamefont {Jacob},
  \citenamefont {Wunsch},\ and\ \citenamefont {Pfannkuche}}]{Jacob2004}%
  \BibitemOpen
  \bibfield  {author} {\bibinfo {author} {\bibfnamefont {D.}~\bibnamefont
  {Jacob}}, \bibinfo {author} {\bibfnamefont {B.}~\bibnamefont {Wunsch}},\ and\
  \bibinfo {author} {\bibfnamefont {D.}~\bibnamefont {Pfannkuche}},\ }\bibfield
   {title} {\bibinfo {title} {Charge localization and isospin blockade in
  vertical double quantum dots},\ }\href
  {https://doi.org/10.1103/PhysRevB.70.081314} {\bibfield  {journal} {\bibinfo
  {journal} {Phys. Rev. B}\ }\textbf {\bibinfo {volume} {70}},\ \bibinfo
  {pages} {081314} (\bibinfo {year} {2004})}\BibitemShut {NoStop}%
\bibitem [{\citenamefont {Silva}\ \emph {et~al.}(2007)\citenamefont {Silva},
  \citenamefont {Villas-Bôas},\ and\ \citenamefont {Ulloa}}]{DiasDaSilva2007}%
  \BibitemOpen
  \bibfield  {author} {\bibinfo {author} {\bibfnamefont {L.~G. D.~D.}\
  \bibnamefont {Silva}}, \bibinfo {author} {\bibfnamefont {J.~M.}\ \bibnamefont
  {Villas-Bôas}},\ and\ \bibinfo {author} {\bibfnamefont {S.~E.}\ \bibnamefont
  {Ulloa}},\ }\bibfield  {title} {\bibinfo {title} {Tunneling and optical
  control in quantum ring molecules},\ }\href
  {https://journals.aps.org/prb/abstract/10.1103/PhysRevB.76.155306} {\bibfield
   {journal} {\bibinfo  {journal} {Physical Review B - Condensed Matter and
  Materials Physics}\ }\textbf {\bibinfo {volume} {76}},\ \bibinfo {pages}
  {155306} (\bibinfo {year} {2007})}\BibitemShut {NoStop}%
\bibitem [{\citenamefont {Climente}(2008)}]{Climente2008}%
  \BibitemOpen
  \bibfield  {author} {\bibinfo {author} {\bibfnamefont {J.~I.}\ \bibnamefont
  {Climente}},\ }\bibfield  {title} {\bibinfo {title} {Tuning the tunnel
  coupling of quantum dot molecules with longitudinal magnetic fields},\
  }\href@noop {} {\bibfield  {journal} {\bibinfo  {journal} {Applied Physics
  Letters}\ }\textbf {\bibinfo {volume} {93}},\ \bibinfo {pages} {223109}
  (\bibinfo {year} {2008})}\BibitemShut {NoStop}%
\bibitem [{\citenamefont {Planelles}\ \emph {et~al.}(2010)\citenamefont
  {Planelles}, \citenamefont {Climente}, \citenamefont {Rajadell},
  \citenamefont {Doty}, \citenamefont {Bracker},\ and\ \citenamefont
  {Gammon}}]{Planelles2010}%
  \BibitemOpen
  \bibfield  {author} {\bibinfo {author} {\bibfnamefont {J.}~\bibnamefont
  {Planelles}}, \bibinfo {author} {\bibfnamefont {J.~I.}\ \bibnamefont
  {Climente}}, \bibinfo {author} {\bibfnamefont {F.}~\bibnamefont {Rajadell}},
  \bibinfo {author} {\bibfnamefont {M.~F.}\ \bibnamefont {Doty}}, \bibinfo
  {author} {\bibfnamefont {A.~S.}\ \bibnamefont {Bracker}},\ and\ \bibinfo
  {author} {\bibfnamefont {D.}~\bibnamefont {Gammon}},\ }\href@noop {}
  {\bibfield  {journal} {\bibinfo  {journal} {Phys. Rev. B}\ }\textbf {\bibinfo
  {volume} {82}},\ \bibinfo {pages} {155307} (\bibinfo {year}
  {2010})}\BibitemShut {NoStop}%
\bibitem [{\citenamefont {Phoenix}\ \emph {et~al.}(2022)\citenamefont
  {Phoenix}, \citenamefont {Korkusinski}, \citenamefont {Dalacu}, \citenamefont
  {Poole}, \citenamefont {Zawadzki}, \citenamefont {Studenikin}, \citenamefont
  {Williams}, \citenamefont {Sachrajda},\ and\ \citenamefont
  {Gaudreau}}]{Phoenix2022}%
  \BibitemOpen
  \bibfield  {author} {\bibinfo {author} {\bibfnamefont {J.}~\bibnamefont
  {Phoenix}}, \bibinfo {author} {\bibfnamefont {M.}~\bibnamefont
  {Korkusinski}}, \bibinfo {author} {\bibfnamefont {D.}~\bibnamefont {Dalacu}},
  \bibinfo {author} {\bibfnamefont {P.~J.}\ \bibnamefont {Poole}}, \bibinfo
  {author} {\bibfnamefont {P.}~\bibnamefont {Zawadzki}}, \bibinfo {author}
  {\bibfnamefont {S.}~\bibnamefont {Studenikin}}, \bibinfo {author}
  {\bibfnamefont {R.~L.}\ \bibnamefont {Williams}}, \bibinfo {author}
  {\bibfnamefont {A.~S.}\ \bibnamefont {Sachrajda}},\ and\ \bibinfo {author}
  {\bibfnamefont {L.}~\bibnamefont {Gaudreau}},\ }\bibfield  {title} {\bibinfo
  {title} {{Magnetic tuning of tunnel coupling between {InAsP} double quantum
  dots in InP nanowires}},\ }\href {https://doi.org/10.1038/s41598-022-08548-8}
  {\bibfield  {journal} {\bibinfo  {journal} {Scientific Reports}\ }\textbf
  {\bibinfo {volume} {12}},\ \bibinfo {pages} {5100} (\bibinfo {year}
  {2022})}\BibitemShut {NoStop}%
\bibitem [{\citenamefont {Krenner}\ \emph {et~al.}(2005)\citenamefont
  {Krenner}, \citenamefont {Sabathil}, \citenamefont {Clark}, \citenamefont
  {Kress}, \citenamefont {Schuh}, \citenamefont {Bichler}, \citenamefont
  {Abstreiter},\ and\ \citenamefont {Finley}}]{Krenner2005}%
  \BibitemOpen
  \bibfield  {author} {\bibinfo {author} {\bibfnamefont {H.~J.}\ \bibnamefont
  {Krenner}}, \bibinfo {author} {\bibfnamefont {M.}~\bibnamefont {Sabathil}},
  \bibinfo {author} {\bibfnamefont {E.~C.}\ \bibnamefont {Clark}}, \bibinfo
  {author} {\bibfnamefont {A.}~\bibnamefont {Kress}}, \bibinfo {author}
  {\bibfnamefont {D.}~\bibnamefont {Schuh}}, \bibinfo {author} {\bibfnamefont
  {M.}~\bibnamefont {Bichler}}, \bibinfo {author} {\bibfnamefont
  {G.}~\bibnamefont {Abstreiter}},\ and\ \bibinfo {author} {\bibfnamefont
  {J.~J.}\ \bibnamefont {Finley}},\ }\bibfield  {title} {\bibinfo {title}
  {Direct observation of controlled coupling in an individual quantum dot
  molecule},\ }\href {https://link.aps.org/doi/10.1103/PhysRevLett.94.057402}
  {\bibfield  {journal} {\bibinfo  {journal} {Physical Review Letters}\
  }\textbf {\bibinfo {volume} {94}},\ \bibinfo {pages} {057402} (\bibinfo
  {year} {2005})}\BibitemShut {NoStop}%
\bibitem [{\citenamefont {Bopp}\ \emph {et~al.}(2022)\citenamefont {Bopp},
  \citenamefont {Rojas}, \citenamefont {Revenga}, \citenamefont {Riedl},
  \citenamefont {Sbresny}, \citenamefont {Boos}, \citenamefont {Simmet},
  \citenamefont {Ahmadi}, \citenamefont {Gershoni}, \citenamefont {Kasprzak},
  \citenamefont {Ludwig}, \citenamefont {Reitzenstein}, \citenamefont {Wieck},
  \citenamefont {Reuter}, \citenamefont {M\"uller},\ and\ \citenamefont
  {Finley}}]{Bopp2022}%
  \BibitemOpen
  \bibfield  {author} {\bibinfo {author} {\bibfnamefont {F.}~\bibnamefont
  {Bopp}}, \bibinfo {author} {\bibfnamefont {J.}~\bibnamefont {Rojas}},
  \bibinfo {author} {\bibfnamefont {N.}~\bibnamefont {Revenga}}, \bibinfo
  {author} {\bibfnamefont {H.}~\bibnamefont {Riedl}}, \bibinfo {author}
  {\bibfnamefont {F.}~\bibnamefont {Sbresny}}, \bibinfo {author} {\bibfnamefont
  {K.}~\bibnamefont {Boos}}, \bibinfo {author} {\bibfnamefont {T.}~\bibnamefont
  {Simmet}}, \bibinfo {author} {\bibfnamefont {A.}~\bibnamefont {Ahmadi}},
  \bibinfo {author} {\bibfnamefont {D.}~\bibnamefont {Gershoni}}, \bibinfo
  {author} {\bibfnamefont {J.}~\bibnamefont {Kasprzak}}, \bibinfo {author}
  {\bibfnamefont {A.}~\bibnamefont {Ludwig}}, \bibinfo {author} {\bibfnamefont
  {S.}~\bibnamefont {Reitzenstein}}, \bibinfo {author} {\bibfnamefont
  {A.}~\bibnamefont {Wieck}}, \bibinfo {author} {\bibfnamefont
  {D.}~\bibnamefont {Reuter}}, \bibinfo {author} {\bibfnamefont
  {K.}~\bibnamefont {M\"uller}},\ and\ \bibinfo {author} {\bibfnamefont
  {J.~J.}\ \bibnamefont {Finley}},\ }\bibfield  {title} {\bibinfo {title}
  {Quantum dot molecule devices with optical control of charge status and
  electronic control of coupling},\ }\href@noop {} {\bibfield  {journal}
  {\bibinfo  {journal} {Advanced Quantum Technologies}\ }\textbf {\bibinfo
  {volume} {5}},\ \bibinfo {pages} {2200049} (\bibinfo {year}
  {2022})}\BibitemShut {NoStop}%
\bibitem [{\citenamefont {Fry}\ \emph {et~al.}(2000)\citenamefont {Fry},
  \citenamefont {Finley}, \citenamefont {Wilson}, \citenamefont {Lemaître},
  \citenamefont {Mowbray}, \citenamefont {Skolnick}, \citenamefont {Hopkinson},
  \citenamefont {Hill},\ and\ \citenamefont {Clark}}]{Fry2000a}%
  \BibitemOpen
  \bibfield  {author} {\bibinfo {author} {\bibfnamefont {P.~W.}\ \bibnamefont
  {Fry}}, \bibinfo {author} {\bibfnamefont {J.~J.}\ \bibnamefont {Finley}},
  \bibinfo {author} {\bibfnamefont {L.~R.}\ \bibnamefont {Wilson}}, \bibinfo
  {author} {\bibfnamefont {A.}~\bibnamefont {Lemaître}}, \bibinfo {author}
  {\bibfnamefont {D.~J.}\ \bibnamefont {Mowbray}}, \bibinfo {author}
  {\bibfnamefont {M.~S.}\ \bibnamefont {Skolnick}}, \bibinfo {author}
  {\bibfnamefont {M.}~\bibnamefont {Hopkinson}}, \bibinfo {author}
  {\bibfnamefont {G.}~\bibnamefont {Hill}},\ and\ \bibinfo {author}
  {\bibfnamefont {J.~C.}\ \bibnamefont {Clark}},\ }\bibfield  {title} {\bibinfo
  {title} {Electric-field-dependent carrier capture and escape in
  self-assembled {InAs/GaAs} quantum dots},\ }\href@noop {} {\bibfield
  {journal} {\bibinfo  {journal} {Applied Physics Letters}\ }\textbf {\bibinfo
  {volume} {77}},\ \bibinfo {pages} {4344} (\bibinfo {year}
  {2000})}\BibitemShut {NoStop}%
\bibitem [{Sup()}]{Supp}%
  \BibitemOpen
  \href@noop {} {\bibinfo {title} {See supplemental material at [url will be
  inserted by publisher] for further details on the sample structure and
  theoretical models}}\BibitemShut {NoStop}%
\bibitem [{\citenamefont {Bayer}\ \emph {et~al.}(1998)\citenamefont {Bayer},
  \citenamefont {Walck},\ and\ \citenamefont {Reinecke}}]{Bayer1998}%
  \BibitemOpen
  \bibfield  {author} {\bibinfo {author} {\bibfnamefont {M.}~\bibnamefont
  {Bayer}}, \bibinfo {author} {\bibfnamefont {S.}~\bibnamefont {Walck}},\ and\
  \bibinfo {author} {\bibfnamefont {T.}~\bibnamefont {Reinecke}},\ }\bibfield
  {title} {\bibinfo {title} {Exciton binding energies and diamagnetic shifts in
  semiconductor quantum wires and quantum dots},\ }\href@noop {} {\bibfield
  {journal} {\bibinfo  {journal} {Physical Review B - Condensed Matter and
  Materials Physics}\ }\textbf {\bibinfo {volume} {57}},\ \bibinfo {pages}
  {6584} (\bibinfo {year} {1998})}\BibitemShut {NoStop}%
\bibitem [{\citenamefont {Bayer}\ \emph {et~al.}(2000)\citenamefont {Bayer},
  \citenamefont {Stern}, \citenamefont {Kuther},\ and\ \citenamefont
  {Forchel}}]{Bayer2000}%
  \BibitemOpen
  \bibfield  {author} {\bibinfo {author} {\bibfnamefont {M.}~\bibnamefont
  {Bayer}}, \bibinfo {author} {\bibfnamefont {O.}~\bibnamefont {Stern}},
  \bibinfo {author} {\bibfnamefont {A.}~\bibnamefont {Kuther}},\ and\ \bibinfo
  {author} {\bibfnamefont {A.}~\bibnamefont {Forchel}},\ }\bibfield  {title}
  {\bibinfo {title} {Spectroscopic study of dark excitons in
  {In}$_x${Ga}$_{1-x}${As} self-assembled quantum dots by a
  magnetic-field-induced symmetry breaking},\ }\href
  {https://doi.org/10.1103/PhysRevB.61.7273} {\bibfield  {journal} {\bibinfo
  {journal} {Phys. Rev. B}\ }\textbf {\bibinfo {volume} {61}},\ \bibinfo
  {pages} {7273} (\bibinfo {year} {2000})}\BibitemShut {NoStop}%
\bibitem [{\citenamefont {Bayer}\ \emph {et~al.}(2002)\citenamefont {Bayer},
  \citenamefont {Ortner}, \citenamefont {Stern}, \citenamefont {Kuther},
  \citenamefont {Gorbunov}, \citenamefont {Forchel}, \citenamefont {Hawrylak},
  \citenamefont {Fafard}, \citenamefont {Hinzer}, \citenamefont {Reinecke},
  \citenamefont {Walck}, \citenamefont {Reithmaier}, \citenamefont {Klopf},\
  and\ \citenamefont {Sch\"afer}}]{Bayer2002}%
  \BibitemOpen
  \bibfield  {author} {\bibinfo {author} {\bibfnamefont {M.}~\bibnamefont
  {Bayer}}, \bibinfo {author} {\bibfnamefont {G.}~\bibnamefont {Ortner}},
  \bibinfo {author} {\bibfnamefont {O.}~\bibnamefont {Stern}}, \bibinfo
  {author} {\bibfnamefont {A.}~\bibnamefont {Kuther}}, \bibinfo {author}
  {\bibfnamefont {A.~A.}\ \bibnamefont {Gorbunov}}, \bibinfo {author}
  {\bibfnamefont {A.}~\bibnamefont {Forchel}}, \bibinfo {author} {\bibfnamefont
  {P.}~\bibnamefont {Hawrylak}}, \bibinfo {author} {\bibfnamefont
  {S.}~\bibnamefont {Fafard}}, \bibinfo {author} {\bibfnamefont
  {K.}~\bibnamefont {Hinzer}}, \bibinfo {author} {\bibfnamefont {T.~L.}\
  \bibnamefont {Reinecke}}, \bibinfo {author} {\bibfnamefont {S.~N.}\
  \bibnamefont {Walck}}, \bibinfo {author} {\bibfnamefont {J.~P.}\ \bibnamefont
  {Reithmaier}}, \bibinfo {author} {\bibfnamefont {F.}~\bibnamefont {Klopf}},\
  and\ \bibinfo {author} {\bibfnamefont {F.}~\bibnamefont {Sch\"afer}},\
  }\bibfield  {title} {\bibinfo {title} {Fine structure of neutral and charged
  excitons in self-assembled {In(Ga)As/(Al)GaAs} quantum dots},\ }\href@noop {}
  {\bibfield  {journal} {\bibinfo  {journal} {Physical Review B - Condensed
  Matter and Materials Physics}\ }\textbf {\bibinfo {volume} {65}},\ \bibinfo
  {pages} {1953151} (\bibinfo {year} {2002})}\BibitemShut {NoStop}%
\bibitem [{\citenamefont {Jacak}\ \emph {et~al.}(2013)\citenamefont {Jacak},
  \citenamefont {Hawrylak},\ and\ \citenamefont {Wojs}}]{jacak2013quantum}%
  \BibitemOpen
  \bibfield  {author} {\bibinfo {author} {\bibfnamefont {L.}~\bibnamefont
  {Jacak}}, \bibinfo {author} {\bibfnamefont {P.}~\bibnamefont {Hawrylak}},\
  and\ \bibinfo {author} {\bibfnamefont {A.}~\bibnamefont {Wojs}},\ }\href@noop
  {} {\emph {\bibinfo {title} {Quantum dots}}}\ (\bibinfo  {publisher}
  {Springer Science \& Business Media},\ \bibinfo {year} {2013})\BibitemShut
  {NoStop}%
\bibitem [{\citenamefont {Wasilewski}\ \emph {et~al.}(1999)\citenamefont
  {Wasilewski}, \citenamefont {Fafard},\ and\ \citenamefont
  {McCaffrey}}]{Wasilewski1999}%
  \BibitemOpen
  \bibfield  {author} {\bibinfo {author} {\bibfnamefont {Z.}~\bibnamefont
  {Wasilewski}}, \bibinfo {author} {\bibfnamefont {S.}~\bibnamefont {Fafard}},\
  and\ \bibinfo {author} {\bibfnamefont {J.}~\bibnamefont {McCaffrey}},\
  }\bibfield  {title} {\bibinfo {title} {{Size and shape engineering of
  vertically stacked self-assembled quantum dots}},\ }\href
  {https://doi.org/10.1016/S0022-0248(98)01539-5} {\bibfield  {journal}
  {\bibinfo  {journal} {Journal of Crystal Growth}\ }\textbf {\bibinfo {volume}
  {201-202}},\ \bibinfo {pages} {1131} (\bibinfo {year} {1999})}\BibitemShut
  {NoStop}%
\bibitem [{\citenamefont {Colombelli}\ \emph {et~al.}(2000)\citenamefont
  {Colombelli}, \citenamefont {Piazza}, \citenamefont {Badolato}, \citenamefont
  {Lazzarino}, \citenamefont {Beltram}, \citenamefont {Schoenfeld},\ and\
  \citenamefont {Petroff}}]{Colombelli2000}%
  \BibitemOpen
  \bibfield  {author} {\bibinfo {author} {\bibfnamefont {R.}~\bibnamefont
  {Colombelli}}, \bibinfo {author} {\bibfnamefont {V.}~\bibnamefont {Piazza}},
  \bibinfo {author} {\bibfnamefont {A.}~\bibnamefont {Badolato}}, \bibinfo
  {author} {\bibfnamefont {M.}~\bibnamefont {Lazzarino}}, \bibinfo {author}
  {\bibfnamefont {F.}~\bibnamefont {Beltram}}, \bibinfo {author} {\bibfnamefont
  {W.}~\bibnamefont {Schoenfeld}},\ and\ \bibinfo {author} {\bibfnamefont
  {P.}~\bibnamefont {Petroff}},\ }\bibfield  {title} {\bibinfo {title}
  {Conduction-band offset of single inas monolayers on gaas},\ }\href@noop {}
  {\bibfield  {journal} {\bibinfo  {journal} {Applied Physics Letters}\
  }\textbf {\bibinfo {volume} {76}},\ \bibinfo {pages} {1146} (\bibinfo {year}
  {2000})}\BibitemShut {NoStop}%
\end{thebibliography}%


\begin{thebibliography}{4}%
\makeatletter
\providecommand \@ifxundefined [1]{%
 \@ifx{#1\undefined}
}%
\providecommand \@ifnum [1]{%
 \ifnum #1\expandafter \@firstoftwo
 \else \expandafter \@secondoftwo
 \fi
}%
\providecommand \@ifx [1]{%
 \ifx #1\expandafter \@firstoftwo
 \else \expandafter \@secondoftwo
 \fi
}%
\providecommand \natexlab [1]{#1}%
\providecommand \enquote  [1]{``#1''}%
\providecommand \bibnamefont  [1]{#1}%
\providecommand \bibfnamefont [1]{#1}%
\providecommand \citenamefont [1]{#1}%
\providecommand \href@noop [0]{\@secondoftwo}%
\providecommand \href [0]{\begingroup \@sanitize@url \@href}%
\providecommand \@href[1]{\@@startlink{#1}\@@href}%
\providecommand \@@href[1]{\endgroup#1\@@endlink}%
\providecommand \@sanitize@url [0]{\catcode `\\12\catcode `\$12\catcode
  `\&12\catcode `\#12\catcode `\^12\catcode `\_12\catcode `\%12\relax}%
\providecommand \@@startlink[1]{}%
\providecommand \@@endlink[0]{}%
\providecommand \url  [0]{\begingroup\@sanitize@url \@url }%
\providecommand \@url [1]{\endgroup\@href {#1}{\urlprefix }}%
\providecommand \urlprefix  [0]{URL }%
\providecommand \Eprint [0]{\href }%
\providecommand \doibase [0]{https://doi.org/}%
\providecommand \selectlanguage [0]{\@gobble}%
\providecommand \bibinfo  [0]{\@secondoftwo}%
\providecommand \bibfield  [0]{\@secondoftwo}%
\providecommand \translation [1]{[#1]}%
\providecommand \BibitemOpen [0]{}%
\providecommand \bibitemStop [0]{}%
\providecommand \bibitemNoStop [0]{.\EOS\space}%
\providecommand \EOS [0]{\spacefactor3000\relax}%
\providecommand \BibitemShut  [1]{\csname bibitem#1\endcsname}%
\let\auto@bib@innerbib\@empty
\bibitem [{\citenamefont {Wasilewski}\ \emph {et~al.}(1999)\citenamefont
  {Wasilewski}, \citenamefont {Fafard},\ and\ \citenamefont
  {McCaffrey}}]{Wasilewski1999}%
  \BibitemOpen
  \bibfield  {author} {\bibinfo {author} {\bibfnamefont {Z.}~\bibnamefont
  {Wasilewski}}, \bibinfo {author} {\bibfnamefont {S.}~\bibnamefont {Fafard}},\
  and\ \bibinfo {author} {\bibfnamefont {J.}~\bibnamefont {McCaffrey}},\
  }\bibfield  {title} {\bibinfo {title} {{Size and shape engineering of
  vertically stacked self-assembled quantum dots}},\ }\href@noop {} {\bibfield
  {journal} {\bibinfo  {journal} {Journal of Crystal Growth}\ }\textbf
  {\bibinfo {volume} {201-202}},\ \bibinfo {pages} {1131} (\bibinfo {year}
  {1999})}\BibitemShut {NoStop}%
\bibitem [{\citenamefont {Schall}\ \emph {et~al.}(2021)\citenamefont {Schall},
  \citenamefont {Deconinck}, \citenamefont {Bart}, \citenamefont {Florian},
  \citenamefont {Helversen}, \citenamefont {Dangel}, \citenamefont {Schmidt},
  \citenamefont {Bremer}, \citenamefont {Bopp}, \citenamefont {H\"ullen},
  \citenamefont {Gies}, \citenamefont {Reuter}, \citenamefont {Wieck},
  \citenamefont {Rodt}, \citenamefont {Finley}, \citenamefont {Jahnke},
  \citenamefont {Ludwig},\ and\ \citenamefont {Reitzenstein}}]{Schall2021}%
  \BibitemOpen
  \bibfield  {author} {\bibinfo {author} {\bibfnamefont {J.}~\bibnamefont
  {Schall}}, \bibinfo {author} {\bibfnamefont {M.}~\bibnamefont {Deconinck}},
  \bibinfo {author} {\bibfnamefont {N.}~\bibnamefont {Bart}}, \bibinfo {author}
  {\bibfnamefont {M.}~\bibnamefont {Florian}}, \bibinfo {author} {\bibfnamefont
  {M.}~\bibnamefont {Helversen}}, \bibinfo {author} {\bibfnamefont
  {C.}~\bibnamefont {Dangel}}, \bibinfo {author} {\bibfnamefont
  {R.}~\bibnamefont {Schmidt}}, \bibinfo {author} {\bibfnamefont
  {L.}~\bibnamefont {Bremer}}, \bibinfo {author} {\bibfnamefont
  {F.}~\bibnamefont {Bopp}}, \bibinfo {author} {\bibfnamefont {I.}~\bibnamefont
  {H\"ullen}}, \bibinfo {author} {\bibfnamefont {C.}~\bibnamefont {Gies}},
  \bibinfo {author} {\bibfnamefont {D.}~\bibnamefont {Reuter}}, \bibinfo
  {author} {\bibfnamefont {A.~D.}\ \bibnamefont {Wieck}}, \bibinfo {author}
  {\bibfnamefont {S.}~\bibnamefont {Rodt}}, \bibinfo {author} {\bibfnamefont
  {J.~J.}\ \bibnamefont {Finley}}, \bibinfo {author} {\bibfnamefont
  {F.}~\bibnamefont {Jahnke}}, \bibinfo {author} {\bibfnamefont
  {A.}~\bibnamefont {Ludwig}},\ and\ \bibinfo {author} {\bibfnamefont
  {S.}~\bibnamefont {Reitzenstein}},\ }\bibfield  {title} {\bibinfo {title}
  {Bright electrically controllable quantum‐dot‐molecule devices fabricated
  by in situ electron‐beam lithography},\ }\href@noop {} {\bibfield
  {journal} {\bibinfo  {journal} {Advanced Quantum Technologies}\ }\textbf
  {\bibinfo {volume} {4}},\ \bibinfo {pages} {2100002} (\bibinfo {year}
  {2021})}\BibitemShut {NoStop}%
\bibitem [{\citenamefont {Pomplun}\ \emph {et~al.}(2007)\citenamefont
  {Pomplun}, \citenamefont {Burger}, \citenamefont {Zschiedrich},\ and\
  \citenamefont {Schmidt}}]{pomplun2007}%
  \BibitemOpen
  \bibfield  {author} {\bibinfo {author} {\bibfnamefont {J.}~\bibnamefont
  {Pomplun}}, \bibinfo {author} {\bibfnamefont {S.}~\bibnamefont {Burger}},
  \bibinfo {author} {\bibfnamefont {L.}~\bibnamefont {Zschiedrich}},\ and\
  \bibinfo {author} {\bibfnamefont {F.}~\bibnamefont {Schmidt}},\ }\bibfield
  {title} {\bibinfo {title} {Adaptive finite element method for simulation of
  optical nano structures},\ }\href@noop {} {\bibfield  {journal} {\bibinfo
  {journal} {physica status solidi (b)}\ }\textbf {\bibinfo {volume} {244}},\
  \bibinfo {pages} {3419} (\bibinfo {year} {2007})}\BibitemShut {NoStop}%
\bibitem [{\citenamefont {Bouarissa}\ and\ \citenamefont
  {Aourag}(1999)}]{Bouarissa1999}%
  \BibitemOpen
  \bibfield  {author} {\bibinfo {author} {\bibfnamefont {N.}~\bibnamefont
  {Bouarissa}}\ and\ \bibinfo {author} {\bibfnamefont {H.}~\bibnamefont
  {Aourag}},\ }\bibfield  {title} {\bibinfo {title} {Effective masses of
  electrons and heavy holes in {InAs, InSb, GaSb, GaAs} and some of their
  ternary compounds},\ }\href
  {https://doi.org/https://doi.org/10.1016/S1350-4495(99)00020-1} {\bibfield
  {journal} {\bibinfo  {journal} {Infrared Physics \& Technology}\ }\textbf
  {\bibinfo {volume} {40}},\ \bibinfo {pages} {343} (\bibinfo {year}
  {1999})}\BibitemShut {NoStop}%
\end{thebibliography}%


%

\end{document}



\title{Magnetic tuning of the tunnel coupling in an optical active quantum dot molecule\\
Supplemental Material}

\author{Frederik Bopp}
\altaffiliation{These authors contributed equally to this work}

\affiliation{%
 Walter Schottky Institut, School of Natural Sciences, and MCQST, Technische Universit\"at M\"unchen, Am Coulombwall 4, 85748 Garching, Germany
}%

\author{Charlotte Cullip}
\altaffiliation{These authors contributed equally to this work}
\affiliation{%
 Walter Schottky Institut, School of Natural Sciences, and MCQST, Technische Universit\"at M\"unchen, Am Coulombwall 4, 85748 Garching, Germany
}%

\author{Christopher Thalacker}
\affiliation{%
 Walter Schottky Institut, School of Natural Sciences, and MCQST, Technische Universit\"at M\"unchen, Am Coulombwall 4, 85748 Garching, Germany
}%

\author{Michelle Lienhart}
\affiliation{%
 Walter Schottky Institut, School of Natural Sciences, and MCQST, Technische Universit\"at M\"unchen, Am Coulombwall 4, 85748 Garching, Germany
}%

\author{Johannes Schall}
\affiliation{%
 Institut f\"ur Festk\"orperphysik, Technische Universit\"at Berlin, Hardenbergstraße 36, 10623 Berlin, Germany
}%

\author{Nikolai Bart}
\affiliation{%
 Lehrstuhl f\"ur Angewandte Festk\"orperphysik, Ruhr-Universit\"at Bochum, Universit\"atsstraße 150, 44801 Bochum, Germany
}%

\author{Friedrich Sbresny}
\affiliation{%
 Walter Schottky Institut, School of Computation, Information and Technology, and MCQST, Technische Universit\"at M\"unchen, Am Coulombwall 4, 85748 Garching, Germany
}%

\author{Katarina Boos}
\affiliation{%
 Walter Schottky Institut, School of Computation, Information and Technology, and MCQST, Technische Universit\"at M\"unchen, Am Coulombwall 4, 85748 Garching, Germany
}%

\author{Sven Rodt}
\affiliation{%
 Institut f\"ur Festk\"orperphysik, Technische Universit\"at Berlin, Hardenbergstraße 36, 10623 Berlin, Germany
}%

\author{Dirk Reuter}
\affiliation{%
Paderborn University, Department of Physics, Warburger Straße 100, 33098 Paderborn, Germany
}%

\author{Arne Ludwig}
\affiliation{%
 Lehrstuhl f\"ur Angewandte Festk\"orperphysik, Ruhr-Universit\"at Bochum, Universit\"atsstraße 150, 44801 Bochum, Germany
}%

\author{Andreas D. Wieck}
\affiliation{%
 Lehrstuhl f\"ur Angewandte Festk\"orperphysik, Ruhr-Universit\"at Bochum, Universit\"atsstraße 150, 44801 Bochum, Germany
}%

\author{Stephan Reitzenstein}
\affiliation{%
 Institut f\"ur Festk\"orperphysik, Technische Universit\"at Berlin, Hardenbergstraße 36, 10623 Berlin, Germany
}%

\author{Filippo Troiani}
\affiliation{%
 Centro S3, CNR-Istituto Nanoscienze, Via Campi 213/a, 41125 Modena, Italy
}%

\author{Guido Goldoni}
\affiliation{%
Centro S3, CNR-Istituto Nanoscienze, Via Campi 213/a, 41125 Modena, Italy
}%
\affiliation{%
Dipartimento di Scienze Fisiche, Informatiche e Matematiche, Universit\`a di Modena e Reggio Emilia, Via Campi 213/a, 41125 Modena, Italy
}%

\author{Elisa Molinari}
\affiliation{%
Centro S3, CNR-Istituto Nanoscienze, Via Campi 213/a, 41125 Modena, Italy
}%
\affiliation{%
Dipartimento di Scienze Fisiche, Informatiche e Matematiche, Universit\`a di Modena e Reggio Emilia, Via Campi 213/a, 41125 Modena, Italy
}%

\author{Kai M\"uller}
\affiliation{%
 Walter Schottky Institut, School of Computation, Information and Technology, and MCQST, Technische Universit\"at M\"unchen, Am Coulombwall 4, 85748 Garching, Germany
}%

\author{Jonathan J. Finley}%
 \email{finley@wsi.tum.de}
\affiliation{%
 Walter Schottky Institut, School of Natural Sciences, and MCQST, Technische Universit\"at M\"unchen, Am Coulombwall 4, 85748 Garching, Germany
}%

\date{\today}

\maketitle
\onecolumngrid
\renewcommand{\figurename}{Fig.}
\renewcommand{\thefigure}{S\arabic{figure}}

\section{Sample and Setup}
\label{sec:Sample}

\begin{figure}
\includegraphics[scale=0.5]{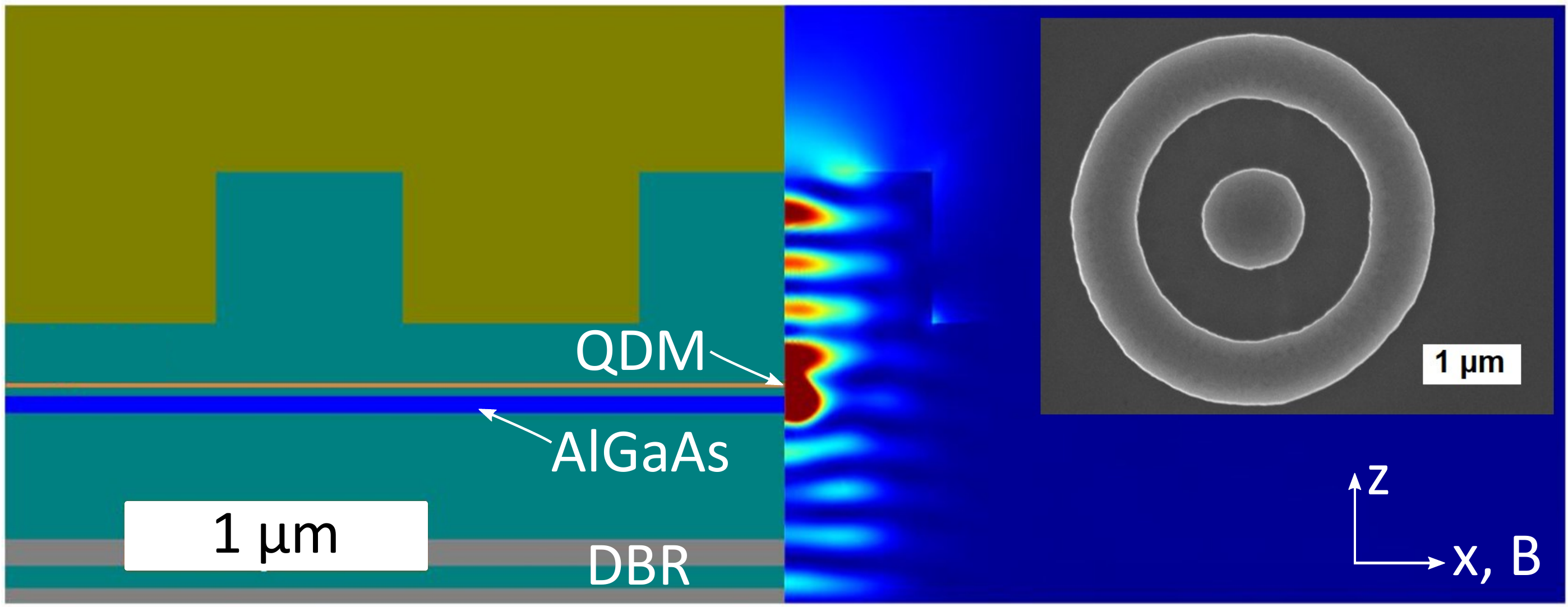}
\caption{\label{fig:SSample} Schematic of the sample design (left) and the simulated electric field distribution (right). A QDM is located at the interface of the two illustrations. The inset shows a scanning electron microscope image of the QDM-circular Bragg grating device.}
\end{figure}

The investigated QDM was grown by solid-source molecular beam epitaxy. It consisted of two vertically stacked InAs QDs, embedded in a GaAs matrix. The height of the top (bottom) QD was fixed to 2.9\,nm (2.7\,nm) via the In-flush technique during growth\,\cite{Wasilewski1999}. This height configuration facilitates electric field-induced tunnel coupling of orbital states in the conduction band.
A wetting layer to wetting layer separation of 10\,nm and an Al$_x$Ga$_{(x-1)}$As barrier ($x=0.33$) with a thickness of 2.5\,nm placed between the dots determines the coupling strength at 0\,T. A 50\,nm thick Al$_x$Ga$_{(x-1)}$As tunnel barrier ($x=0.33$) was grown 5 nm below the QDM to prolong electron tunneling times. The molecule was embedded into a p-i-n diode, with the doped regions used as contacts to gate the sample. The diode contacts are placed more than 150\,nm from the molecule to prevent uncontrolled charge tunneling into the QDM. Furthermore, a distributed Bragg reflector was grown below the diode and a circular Bragg grating is positioned deterministically via in-situ electron beam lithography above an individual QDM to improve photon in- and outcoupling efficiencies\,\cite{Schall2021}. Figure \ref{fig:SSample} shows a schematic of the sample design (left) and simulated electric field distribution (right) calculated using JCMSuite\,\cite{pomplun2007}. The inset depicts a scanning electron microscope image of the circular Bragg grating.
All measurements are performed at 10 K inside an Oxford Instruments magnet. A tunable diode laser is used to excite the QDM.

\section{Two-state model}
\label{sec: $X^0$ Hamiltonian}

The neutral exciton in our system can be described by the following Hamiltonian:
\begin{align}
    H_X^0 = E_0 \cdot \hat{I} + 
    \begin{bmatrix}
        \Gamma - edF & t \\
        t & 0
    \end{bmatrix} + \hat{H}_{QCSE},
\end{align}
where $E_0$ is the zero-field energy of the direct exciton, $\hat{I}$ is the identity matrix, $\Gamma$ is the energy needed to move an electron from the upper to the lower dot, and $t$ is the tunnel coupling. $edF$ accounts for the Stark shift for inter-dot excitons, with $e$ the electron charge, $d$ the separation between electron and hole, and $F$ the electric field.
The final term $\hat{H}_{QCSE}$ accounts for the energy shift from the quantum-confined Stark-effect, and is given by
\begin{align}
    \Delta E_{QCSE} = \vec{p} \cdot \vec{F} + \beta F^2,
\end{align}
where $\vec{p}$ is the dipole moment and $\beta$ is the polarizability.
Solving the full Hamiltonian gives the following eigenvalues:
\begin{align}
    E_{\pm} = \frac{1}{2}(2E_0-edF+\Gamma + 2\Delta E_{QCSE}\\ 
    \pm \sqrt{(edF-\Gamma)^2 + (2t)^2}).
\end{align}
In the main manuscript, these energies for the symmetric (dotted pink) and anti-symmetric (dotted green) eigenstates are shown in Figure 1 (b).
The splitting $\Delta$E between these eigenenergies is then given as 
\begin{align}
\label{equ:DEanalytical}
    \Delta E = \sqrt{(edF-\Gamma)^2 + (2t)^2},
\end{align}
where $2t$ is the minimum splitting between the two eigenstates, i.e. at the avoided crossing. This equation is used for fitting the data points in Figure 3 (a) of the main manuscript.

\section{Single particle states in 3D confinement potential}
\label{sec:3dconfinment}
To solve Equation 2 of the main manuscript, we discretize the confinement potential on a real-space grid of $N=N_1\times N_2\times N_3$ points, where $N=64\times64\times128$. The simulation space has a size of $700\times700\times220$\,nm$^3$. Each point is identified by the vector $\textbf{r}_i=\sum_{k=1}^3\left(\lambda_i^k-N_k/2\right)\Delta_k\textbf{e}_k$, with $\lambda_i^k=1,...,N_k$ and $\textbf{e}_{1,2,3}=\textbf{x},\textbf{y},\textbf{z}$. By rephrasing the resulting finite-difference equation, we obtain a discrete eigenvalue problem:

\begin{equation}
\begin{split}
\label{equ:dep}
\sum_{j=1}^{N}\biggl\{\frac{1}{2m_\chi^*}\biggl[-\hbar^2\left(\nabla^2\right)_{ij}+\frac{2i\hbar q_\chi}{c}\left(\textbf{A}\cdot\nabla\right)_{ij}
+\frac{q_\chi^2}{c^2}\left(\textbf{A}^2\right)_{ij}\biggr]+V^\chi_{ij}\biggr\}\phi^\chi_{\alpha,j}=\epsilon_\alpha\phi^\chi_{\alpha,i}\ \text{,}
\end{split}
\end{equation}
where $\phi^\chi_{\alpha,i}=\phi^\chi_\alpha(\textbf{r}_i)$ is the eigenstate of the Hamiltonian, $\chi=e,h$ for electrons and holes, and $q_h=-q_e=|e|$. In the real-space basis, the vector- and confining-potential operators are diagonal, thus $V_{ij}^\chi=\delta_{i,j}V_\chi(\textbf{r}_i)$ and $(\textbf{A}^2)_{ij}=\delta_{i,j}[\textbf{A}(\textbf{r}_i)]^2$. When applying the differential operator on the wave function $\phi^\chi_{\alpha,i}$, we obtain: 
\begin{equation}
\begin{split}
\label{equ:dep}
\sum_{j=1}^{N} & \left(\nabla^2\right)_{ij}\phi_{\alpha,j}=\sum_{k=1}^{3}\frac{\phi_\alpha\left(\textbf{r}_i+\Delta_k\textbf{e}_k\right)-2\phi_\alpha\left(\textbf{r}_i\right)+\phi_\alpha\left(\textbf{r}_i-\Delta_k\textbf{e}_k\right)}{\Delta_k^2}
\end{split}
\end{equation}
and
\begin{equation}
\begin{aligned}
\label{equ:dep}
\sum_{j=1}^{N} & \left(\textbf{A}\cdot\nabla\right)_{ij}\phi_{\alpha,j} = \sum_{k=1}^{3}A_{k,i}\frac{\phi_\alpha\left(\textbf{r}_i+\Delta_k\textbf{e}_k\right)-\phi_\alpha\left(\textbf{r}_i-\Delta_k\textbf{e}_k\right)}{2\Delta_k}\ \text{,}
\end{aligned}
\end{equation}where $A_{k,i}=A_k(\textbf{r}_i)$.

The Hamiltonian is calculated for a 3D-potential, which consists of an asymmetric double square well along the \textbf{z} direction and parabolic potentials along the \textbf{x} and \textbf{y} direction. The square wells have a width of $d_1=2.7$\,nm and $d_2=2.9$\,nm, which corresponds to the height of the bottom and top QD, respectively. The potential depth $V_0=690$\,meV matches the conduction band offset between GaAs and InAs. An electric field equivalent to $3.63$\,meV/nm is included into the model to facilitate the hybridisation of the two lowest electron eigenstates. The model is calibrated by setting the effective electron mass $m_e^*$, as the only free parameter, such that the energy gap between the two lowest eigenstates matches the experimental results. We obtain an effective mass of $m^*=0.0495\,m_{e,0}$ ($m_{e,0}$ is the free-electron mass), which is between the recommended effective masses of GaAs ($m^*_{\text{GaAs}}=0.067$) and InAs ($m^*_{\text{InAs}}=0.023$)\,\cite{Bouarissa1999}. A magnetic field is applied along the in-plane direction \textbf{x}.

\begin{figure}
\includegraphics{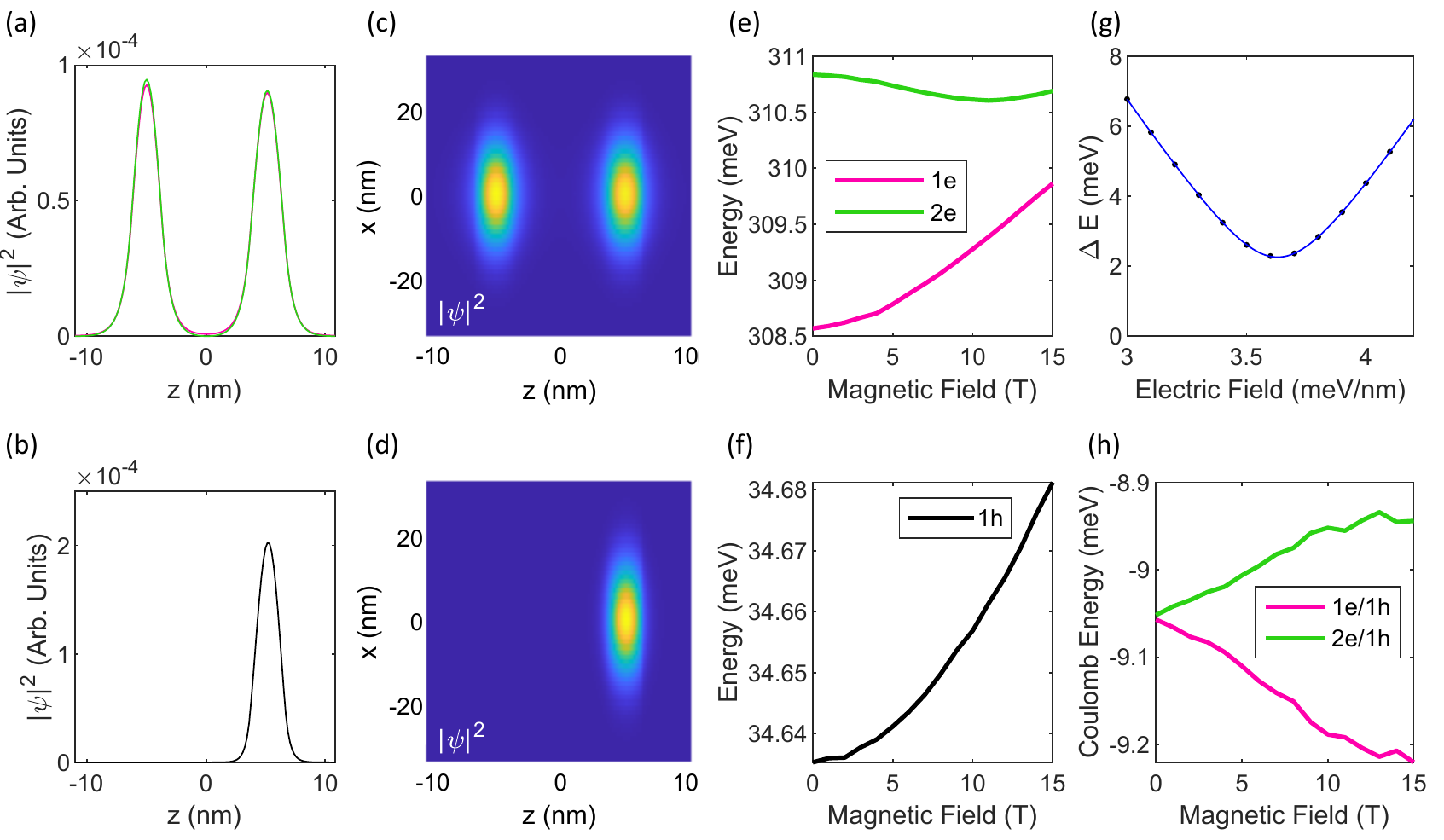}
\caption{\label{fig:Stheo} (a): Absolute value of the wave function along the growth axes of the lowest (pink) and second lowest (green) electron eigenstate at resonance, and of the lowest hole eigenstate (b). (c) and (d): Two dimensional wave function along \textbf{x} and \textbf{z} for the lowest electron and hole eigenstate, respectively. (e) and (f) magnetic field dependent energy of the two lowest electron and lowest hole eigenstate, respectively. (g): Energy difference of the two lowest electron eigenstates for varying bias (black). The data points are fitted using equation 5 (blue) (h) Coulomb energy between the lowest (second lowest) electron and lowest hole eigenstate in pink (green).}
\end{figure}

Figure \ref{fig:Stheo}\,(a) and (b) show the absolute value of the electron and hole wave functions along the growth direction \textbf{z} at $0$\,T. The green (pink) data set represents the lowest (second lowest) electron eigenstate $\phi_1^e$ ($\phi_2^e$). Hereafter we refer to them as e1 and e2, respectively. Panels\,(c) and (d) show surface plots of the wave function for the lowest electron and hole eigenstate, respectively. The magnetic field dependent single particle energies of e1, e2 and the lowest hole eigenstate are shown in panels (e) and (f), respectively. Bias sweeps are performed to ensure that the simulations are performed at the coupling condition. Figure \ref{fig:Stheo}\,(g) shows the energy difference between e1 and e2 as a function of the bias at $0$\,T. The energy difference is fitted by equation \ref{equ:DEanalytical} to obtain the resonance bias and the minimum energy splitting.

In the experiment, we analyze the energy splitting $\Delta$E of the neutral exciton, instead of a single particle state. To account for the Coulomb interaction, we calculate the direct matrix involving e1, e2 and the lowest energy hole state h1. 
The direct and attractive Coulomb matrix elements are given by:
\begin{equation}
\begin{split}
\label{equ:dep}
V_{\alpha\alpha\beta\beta}^{eh}=\iint
\frac{[\phi_\alpha^e(\textbf{r})]^*[\phi_\beta^{h}(\textbf{r'})]^*\phi_\beta^{h}(\textbf{r'})\phi_\alpha^e(\textbf{r})}
{\kappa_r|\textbf{r}-\textbf{r'}|} 
\,d\textbf{r}\,d\textbf{r'}\ \text{,}
\end{split}
\end{equation}
where $\kappa_r$ is the static dielectric constant of the semiconductor medium. The matrix elements are numerically calculated from the expression
\begin{equation}
\begin{split}
\label{equ:dep}
V_{\alpha\beta\gamma\delta}^{\chi\chi'}=
\pm\frac{e^2}{\kappa}\int\mathcal{F}^{-1}\left[\frac{1}{k^2}\tilde{\Phi}_{\alpha\beta}^{\chi}(\textbf{k})\right]\Phi_{\gamma\delta}^{\chi'}(\textbf{r})
\,d\textbf{r}\ \text{,}
\end{split}
\end{equation}
where $\Phi_{\alpha\beta}^{\chi}(\textbf{r})=[\phi_\alpha^\chi(\textbf{r})]^*\phi_\beta^\chi(\textbf{r})$, and $\tilde{\Phi}_{\alpha\beta}^{\chi}(\textbf{k})=\mathcal{F}[\phi_{\alpha\beta}^\chi(\textbf{r})]$ is its Fourier transform.

The exciton energy is then computed by including the Coulomb interaction in first-order perturbation theory:
\begin{equation}
    E_{X_{0,l}} = \epsilon^e_l + \epsilon^h_1 + V_{ll11}^{eh}\ ,
\end{equation}
where $l=1,2$ specifies the involved electron state and the corresponding exciton. 
The energy splitting $\Delta$E plotted in Figure 2\,(b) incorporates the single particle energy difference $\Delta$E$_{SP}$ as well as the difference in Coulomb energy $\Delta\text{E}_{C}$, and is given by: 
\begin{equation}
    \Delta E^{Th} = E_{X_{0,2}} - E_{X_{0,1}} = \epsilon^e_2 - \epsilon^e_1 + V_{2211}^{eh} - V_{1111}^{eh} = \Delta\text{E}_{SP}+\Delta\text{E}_{C}.
\end{equation}

Figure \ref{fig:Stheo}\,(h) shows the Coulomb energy between e1 (e2) and lowest energy hole state in pink (green). 
\bibliography{Supplemental.bbl}